\documentclass[sigconf]{acmart}

\setcopyright{none}
\renewcommand\footnotetextcopyrightpermission[1]{}
\usepackage{booktabs} 

\usepackage[ruled,noend,commentsnumbered]{algorithm2e}
\usepackage{algpseudocode}
\usepackage{graphicx}
\usepackage{color}
\usepackage[all]{nowidow}

\usepackage{color,verbatim,afterpage,url,xfrac}
\graphicspath{{figures/}}

\long\def\comment#1{}

\usepackage{graphicx}
\usepackage{subfig}
\usepackage{subscript}
\usepackage{MnSymbol,wasysym}
\usepackage{listings}
\usepackage{multirow,rotating}
\usepackage{fancyhdr} 
\usepackage{moresize}
\usepackage{alltt}
\usepackage{textgreek}

\date{}

\SetKwProg{MyRoutine}{Procedure}{}{}

\graphicspath{{figures/}}

\usepackage[normalem]{ulem}
\usepackage{enumitem}

\begin{document}


\title{\fontsize{16pt}{16pt}\selectfont Classical Optimizers for Noisy Intermediate-Scale Quantum Devices}

\author{Wim Lavrijsen, Ana Tudor,  Juliane M\"{u}ller, Costin Iancu, Wibe de Jong \\
Lawrence Berkeley National Laboratory \\
 \it \small{\{wlavrijsen,julianemueller,cciancu,wadejong\}@lbl.gov,
anamtudor@berkeley.edu} \\
}

\renewcommand{\shortauthors}{W.~Lavrijsen et. al.}

\begin{abstract}
{\it We present a collection of optimizers tuned for usage on Noisy
Inter-mediate-Scale Quantum (NISQ) devices.
Optimizers have a range of applications in quantum computing, including the
Variational Quantum Eigensolver (VQE) and Quantum Approximate Optimization
(QAOA) algorithms.
They are also used for calibration tasks, hyperparameter tuning, in
machine learning, etc.
We analyze the efficiency and effectiveness of different optimizers in a
VQE case study.
VQE is a hybrid algorithm, with a classical minimizer step driving the next
evaluation on the quantum processor.
While most results to date concentrated on tuning the quantum VQE circuit, we
show that, in the presence of quantum noise, the classical minimizer step
needs to be carefully chosen to obtain correct results.
We explore  state-of-the-art  gradient-free optimizers capable of
handling noisy, black-box, cost functions and stress-test them using a quantum
circuit simulation environment with noise injection capabilities on individual
gates.
Our results indicate that specifically tuned optimizers are crucial to
obtaining valid science results on NISQ hardware, and will likely remain
necessary even for future fault tolerant circuits. }
\end{abstract}

\maketitle

\title{Classical Optimizers for Noisy Intermediate-Scale Quantum Devices}

\section{Introduction}
\label{sec:intro}

Hybrid quantum-classical algorithms are promising candidates to exploit the
potential advantages of quantum computing over classical computing on
current quantum hardware.
Target application domains include the computation of physical and chemical
properties of atoms and molecules~\cite{feynman1982}, as well as optimization
problems~\cite{farhi2014quantum,Wang_2018} such as graph MaxCut.

These hybrid algorithms execute a classical optimizer that iteratively queries
a quantum algorithm that evaluates the optimization objective.
An example is the Variational Quantum Eigensolver (VQE)
algorithm~\cite{McClean2015} applied in chemistry, where the objective
function calculates the expectation value of a Hamiltonian $\mathcal{H}$
given an input configuration of a simulated physical system.
The classical side variationally changes the parametrized input, until
converged on a global minimum, finding the corresponding eigenvalue and
eigenstate.
Since $\mathcal{H}$ describes the energy evolution, this global minimum
represents the ground state energy of the system.
Quantum Approximate Optimization
Algorithms~(QAOA)~\cite{farhi2014quantum,Wang_2018} employ a similar
approach.

For the foreseeable future, quantum algorithms will run on ``Noisy
Intermediate-Scale Quantum" (NISQ) devices, which provide a small number
of noisy, uncorrected qubits.
Hybrid methods are considered auspicious on such devices due to:
\begin{enumerate}[leftmargin=*]
\item reduced chip coherence time requirements because of the single
  Hamiltonian evaluation per circuit execution; and
\item their iterative nature, making them more robust to noise.
\end{enumerate}
These expectations concern only the quantum side of the hybrid approach.
But, as we will show in this paper, understanding the impact of noise on
the classical side is just as important: the performance and mathematical
guarantees on convergence and optimality of commonly used classical
optimizers rests on premises that are negated by noisy objective functions.
Consequently, optimizers may converge too early and miss the global
minimum, halt in noise-induced local minima, or even fail to converge at
all.

For chemistry, the necessity of classical optimizers for VQE that are
robust to hardware noise, has already been recognized~\cite{McClean2015}.
However, the first published hardware studies side-stepped optimizers by
performing a full phase space
exploration~\cite{Li_2017,Temme_2017,Dumitrescu_2018} and backfitting the
solution to zero noise.
This works for low qubit count and few minimization parameters, but is not
tractable at the $\mathcal{O}(100)$ qubit concurrency expected on
NISQ-era devices, nor for realistic problems with many parameters.
To our knowledge, QAOA studies also ignore the effects of the noise on the
classical optimizers.

In this study, we want to understand the requirements on classical
optimizers for hybrid algorithms running on NISQ hardware and which
optimization methods best fulfill them.
We use VQE as a test vehicle, but expect the findings to be readily
applicable to QAOA and other hybrid methods which employ classical numerical
optimization.
The goals and contributions of our empirical study are twofold:
\begin{itemize}[leftmargin=*]
\item A practical software suite of classical optimizers, directly usable
  from Python-based quantum software stacks, together with a tuning guide.
  We consider factors such as the quality of the initial solution and
  availability of bounds, and we test problems with increasing number of
  parameters to understand scalability of the selected methods.
\item A study of the optimizers' sensitivity to different types of noise,
  together with an analysis of the impact on the full VQE algorithm.
  We consider the domain science perspective: some level of experimental
  error is expected and acceptable, as long as the result is accurate and
  the errors can be estimated.
  We run simulations at different noise levels and scale, for several
  science problems with different optimization surfaces, finding the
  breaking points of the minimizers and the algorithm for each.
\end{itemize}

We have taken a very practical tack and first evaluated the minimizers from
SciPy~\cite{scipy_web}.
These include methods such as the quasi-Newton BFGS~\cite{NoceWrig06} 
algorithm, and are the default choice of many practitioners.
Most optimization tools in standard Python and MATLAB software are not
noise-aware and, as we have found in our evaluations, actually fail in the
presence of quantum noise.
Some optimizers are more robust due to the smoothing effect of the underlying
methods used (e.g.\ local modeling in trust region methods), but that is
seldom by explicit design.

Fortunately, applied mathematicians in the optimization community have long
been working on this type of problem and have provided high quality, open
source, software.
Based on their recommendation, our final selection contains representative
methods of (hybrid) mesh (ImFil~\cite{ImFil11}, NOMAD~\cite{nomad}); local
fit (SnobFit~\cite{snobfit}); and trust regions
(PyBobyqa~\cite{pybobyqa1, pybobyqa2}).
Python and C++ are far more widely used in quantum computing than MATLAB. 
Thus, we have rewritten optimizers where necessary from MATLAB into Python,
while ensuring, through a suite of unit tests, reproducible deterministic
behavior after porting, and provided consistent interfaces and plugins for
high level quantum frameworks such as Qiskit~\cite{qiskit} and
Cirq~\cite{cirq}.
These products have been packaged into {\sc scikit-quant}~\cite{urlsciq}.
The optimization package in {\sc scikit-quant} also provides tutorial
notebooks with tips and hints for hyper-parameter optimization, and an
evaluation harness to quickly assess applicability to new problems.

{\sc scikit-quant} has been evaluated on three VQE problems (ethylene
$C_2H_6$ rotation and bond stretching, and Hubbard model simulation), each
with different optimization requirements.
The results indicate that a {\em suite} of minimizers is needed to match
specific strengths to specific problems.
Achieving high quality solutions is aided by domain science information, if
available, such as good initial parameters, knowledge of local minima, or
the need to search around inaccessible regions.
Such information is problem specific and in practice we observe different
performance benefits with different optimizers from its inclusion.
Where this information is {\em not} available, our study indicates that the
best results are obtained by composing local and global optimizers,
leveraging their respective strengths, during the VQE algorithm run.

The organization of this paper is as follows.
In Section~\ref{sec:numopt_background}, we give a brief background on numerical
optimization and our requirements on optimizers.
In Section~\ref{sec:scq} we describe the optimizers available in
{\sc scikit-quant} in more detail.
We provide the necessary background on hybrid quantum-classical algorithms in
Section~\ref{sec:hqca} and we describe the impact of noise in
Section~\ref{sec:noise_impact}.
Our numerical experiments are presented in Section~\ref{sec:results} and
discussed in Section~\ref{sec:discusion}.
We compare our work with related studies in Section~\ref{sec:related} and
finally summarize the main conclusions in Section~\ref{sec:conclusion}.

\section{Numerical Optimization}
\label{sec:numopt_background}

In variational hybrid quantum-classical algorithms, such as VQE, the
execution on the quantum processor evaluates the objective function to be
optimized classically.
We restrict ourselves to derivative-free methods.
Evaluation of, and comparison with, gradient-based and gradient-approximating
methods~\cite{Kubler2020adaptiveoptimizer}\cite{crooks2019gradients}
is worthy of a study of its own.
For a deterministic function
$f:\Omega\subset {\rm I\!R}^n \rightarrow {\rm I\!R}$
over a domain $\Omega$ of interest that has lower and upper bounds on the
problem variables, derivative-free algorithms require only evaluations of
$f$  but no derivative information.
They assume that the derivatives of $f$ are neither symbolically nor
numerically available, and that bounds, such as Lipschitz constants, for the
derivatives of $f$ are also unavailable.

Optimizers are judged on the quality of the solution and on their speed and
scalability.
A good solution has a short distance to the true global optimum, high
accuracy of the optimal parameters found, or both.
A good overview and thorough evaluation of derivative-free algorithms can be
found in Rios et al.~\cite{Rios2013}.
The main criteria for matching an optimizer to a problem are the convexity
and the smoothness of the optimization surfaces.
Convexity has the familiar meaning; smoothness in our context requires
that the function is ``sufficiently often differentiable''.
In VQE, the shape of the optimization surface is determined by the ansatz,
and although typical surfaces are smooth, noise can change this considerably.

\begin{figure*}
\begin{minipage}[l]{1.0\columnwidth}
\centering
\includegraphics[height=2.5in]{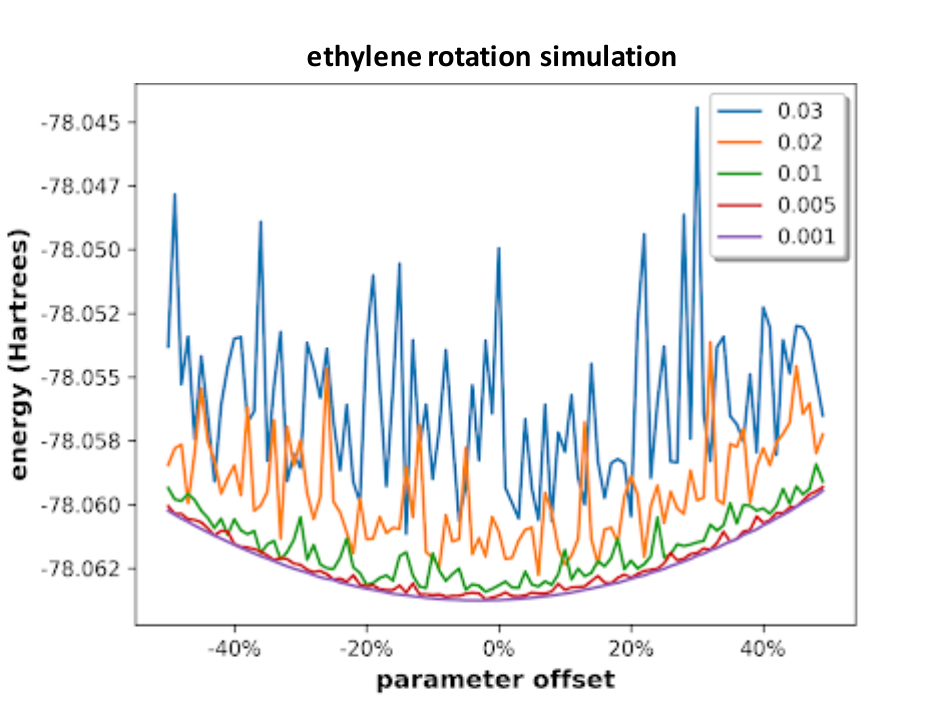}
\caption{\it \small The optimization surface in the main parameter for
  ethylene rotation simulation as a function of Gaussian gate noise.
  The surface transforms from convex and smooth to non-convex, non-smooth.}
\label{fig:ethynoise}
\end{minipage}
\vspace{-0.08in}
\hfill{}
\begin{minipage}[r]{1.0\columnwidth}
\centering
\includegraphics[height=2.5in]{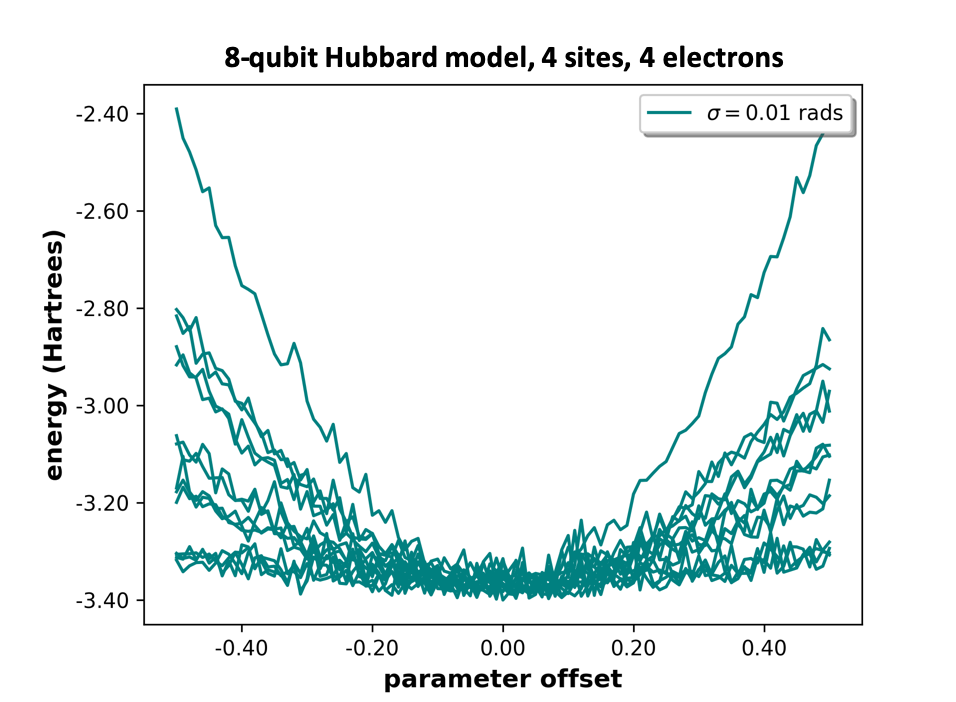}
\caption{\label{fig:hubbardnoise}\it \small Optimization surfaces of
  all 14 parameters with Gaussian gate noise of
  $\sigma = 0.01 \mathrm{rad.}$ in a Hubbard model simulation of 4 sites
  with 4 electrons (see Section~\ref{sec:results} for full details).}
\end{minipage}
\vspace{-0.08in}
\end{figure*}

Figure~\ref{fig:ethynoise} shows the evolution of the optimization surface
for a single parameter in a simple VQE problem (rotation/torsion of an
ethylene molecule; 4 qubits, 2 parameters) for increasing levels of Gaussian
gate noise (detailed background on this and other studies is provided in
Sections~\ref{sec:hqca}~and~\ref{sec:noise_impact}).
For low noise, the optimization surface is convex around the global minimum
and smooth.
For increasing levels of noise, the optimization surface becomes both
non-convex and non-smooth.
It gets substantially worse for more complex problems: because circuit depth
increases, because the number of parameters increases the likelihood of
noise-induced local minima, and because entanglement over many qubits means
that the effects of gate noise become non-local.
This can be seen in Figure~\ref{fig:hubbardnoise}, which displays the effect
of noise on an 8 qubit Hubbard model simulation, with 14 parameters at a
moderate level of gate noise of $\sigma = 0.01 \mathrm{rad.}$ (cf.\ the
mid-range in the ethylene figure).
We are thus interested in optimizers that perform well across the whole range
of behaviors: convex and non-convex surfaces, smooth and non-smooth surfaces.

\subsection{Optimizer Selection Criteria}

The criteria for selecting optimizers that  we considered are:
\begin{enumerate}[leftmargin=*]
\item Ability to find a good solution in the presence of noise, potentially
  using different methods for different types of surfaces and noise impacts.
\item Scalability with the number of parameters, as this determines the
  asymptotic behavior on future quantum hardware that allows the simulation of
  larger problems.
\item Number of samples (queries to the objective function) required and
  precision needed, which affects scaling and wall-clock time spent on the
  quantum chip.
\item Implementation performance and ability to parallelize, as these affect
  scaling and wall-clock time spent on the classical side.
\end{enumerate}

There are two common strategies for optimizing  noisy outcomes: optimize
for the expected value of the response, or for the worst
case~\cite{Powell2014}.
Quantum simulations, being probabilistic in nature, fit the former: many
runs (``shots'') of a circuit are required to obtain the output distribution,
which is then inclusively averaged over local noise sources.

\subsection{Baseline Optimizers}
\label{sec:baseline}

The BFGS implementation from SciPy~\cite{scipy_web} is a common first choice
for optimizations of objective functions for which no derivative can be
calculated.
In this algorithm, first and second derivatives are instead constructed from
evaluations.
Since each new evaluation is added to the current derivative estimate with
equal weight to all points collected so far combined, noisy results will
quickly throw it off-track and the algorithm will thus not perform well in
the presence of noise.
However, because it is so well known and because it is commonly used in other
papers for comparative purposes, we will keep it as one of our baselines as
well.

We have also evaluated a range of other methods for which implementations are
readily available in Python, such as the Nelder-Mead simplex
method~\cite{nelder-mead} (considered by McClean et al.~\cite{McClean2015} in
their initial VQE analysis paper), RBFOpt~\cite{rbfopt}, Cobyla~\cite{cobyla},
DYCORS\cite{dycors}, and CMA-ES~\cite{cmaes,cmapy}.
These methods do not make the hard assumptions about data quality that BFGS
does, leaving them somewhat more robust to noise.
Based on our evaluation, we find Cobyla to outperform all others, and thus we
use it as a second baseline for our comparisons.

\comment{Under the assumption that the objective function is still continuously
differentiable, quasi-Newton methods can be used.
These approximate the first (and often the second) derivative from the
evaluations at different points.
Such methods work better if a detailed understanding of the noise is
available, allowing selection of good step sizes and properly weigh
evaluations when incorporating them into the approximation of the derivatives.
In the case of BFGS, which has been used by VQE developers for algorithm
development on quantum simulators\footnote{As opposed to real hardware.}, each
new evaluation is instead added to the current derivative estimate with equal
weight to all points collected so far combined.
This means that BFGS is easily thrown off when function values are noisy.

Given that it is still a common first choice, we retain BFGS as a baseline for
comparisons for our initial experiments and candidate optimizer selection for
{\sc scikit-quant}.
We use the SciPy~\cite{scipy_web} BFGS implementation and tune it for all input
problems.
We have also evaluated a range of other methods for which implementations are
readily available in Python, such as the Nelder-Mead simplex
method~\cite{nelder-mead} (considered by McClean et al.~\cite{McClean2015} in
their initial VQE analysis paper), RBFOpt~\cite{rbfopt}, Cobyla~\cite{cobyla},
DYCORS\cite{dycors}, and CMA-ES~\cite{cmaes,cmapy}.
These methods do not make the hard assumptions about data quality that BFGS
does, leaving them somewhat more robust to noise.
Based on our evaluation, we find Cobyla to outperform and thus we use it as a
second baseline for subsequent comparisons.}

\comment{A different type of minimizers that focuses on derivative-free methods are
so-called ``blackbox optimizers").
These employ sampling techniques, possibly combined with local models, to
and do not impose any  requirements on the
objective function. For example, the Nelder-Mead simplex method~\cite{nelder-mead} is
considered by McClean et al~\cite{mcclean} in their initial VQE
analysis paper.

There are two common strategies (which can be intermixed within the same
minimizer) sampling based on meshes and based on local fitting.
The former requires that the parameter space that is searched is bounded, the
latter requires some functional description of the local optimization surface.}

\section{{\sc scikit-quant} Optimizers}
\label{sec:scq}

The initial selection of optimizers packaged in {\sc scikit-quant} consists
of NOMAD, ImFil, SnobFit, and BOBYQA; each detailed in the rest of this
section.
This choice is motivated by the evaluation of Rios et al.~\cite{Rios2013}
combined with open-source availability and ease of porting\footnote{Note
  that while we ported the same algorithms, they evaluated different
  implementations, which may affect some of the total running time.}
to Python.
Rios et al.~\cite{Rios2013} indicate the following trends:
\begin{itemize}[leftmargin=*]
\item Scalability: SnobFit and NOMAD may have challenges with the number of
  parameters (tested up to 300).
  ImFil and BOBYQA are among the fastest optimizers.
\item For convex optimization surfaces, BOBYQA and SnobFit perform well
  for smooth surfaces, while NOMAD and ImFil outperform on non-smooth
  surfaces.
\item For non-convex optimization surfaces, SnobFit and NOMAD are good
  for smooth surfaces, while for non-smooth surfaces ImFil and NOMAD are
  recommended.
\end{itemize}

In the rest of this section we give a short description of each algorithm
together with their tunable knobs that affect their performance and solution
quality.
As common characteristics we note that all derivative-free optimizers employ
sampling strategies and require a minimum number of samples to get started.
This allows a common interface to employ parallelization of the quantum step,
even if the original codes do not support this directly.
Sampling requires that the parameter space is bounded, or that search vectors
are provided.
Most optimizers can make use of further detailed science domain information,
such as the magnitude and shape of uncertainties, local functional descriptions,
inaccessible regions, etc.
If no such information is provided or available, they assume reasonable
defaults, e.g.\ homogeneous, symmetric, uncertainties; and qubic or quadratic
local functional behavior on a small enough region.
Inaccessible regions can simply be communicated by returning $\mathrm{NaN}$
from the objective function.

\subsection{NOMAD}
\label{ssec:nomad}

NOMAD, or {\em Nonlinear Optimization by Mesh Adaptive Direct
Search (MADS)}~\cite{nomad} is a C++ implementation of the MADS
algorithm~\cite{abramson2009, audet2006, audet2009}.
MADS searches the parameter space by iteratively generating a new sample 
point from a  mesh that is adaptively adjusted based on the progress 
of the search. If the newly selected sample point does not improve the current 
best point, the mesh is refined. NOMAD uses   two  steps ({\em search} and
{\em poll}) alternately until some preset stopping criterion (such
as minimum mesh size, maximum number of failed consecutive trials, or maximum
number of steps) is met.
The search step can return any point on the current mesh, and therefore offers no
convergence guarantees. 
If the search step fails to find an improved solution, the poll step is used to 
explore  the neighborhood of the current best
solution. The poll step is central to the convergence analysis of NOMAD, and
therefore any hyperparameter optimization or other tuning to make progress should
focus on the poll step.
Options include: poll direction type (local model, random, uniform angles,
etc.), poll size, and number of polling points.

The use of meshes means that the number of evaluations needed scales at least
geometrically with the number of parameters to be optimized.
It is therefore important to restrict  the search space as much as possible
using bounds and, if the science of the problem so indicates, give preference
to polling directions of the more important parameters.

In {\sc scikit-quant} we incorporate the published open-source NOMAD code
through a modified Python interface.

\subsection{ImFil}
\label{ssec:imfil}

Implicit Filtering (ImFil~\cite{ImFil11}) is an algorithm designed for problems
with local minima caused by high-frequency, low-amplitude noise and with an
underlying  large scale structure that is easily optimized.
ImFil uses difference gradients during the search and can be considered as an 
extension  of coordinate search.
In ImFil, the optimization is controlled by evaluating the objective function
at a cluster (or stencil) of points within the given bounds.
The minimum of those evaluations then drives the next cluster of points,
using first-order interpolation to estimate the derivative, and aided by
user-provided exploration directions, if any.
Convergence is reached if the ``budget'' for objective function evaluations is
spent, if the smallest cluster size has been reached, or if incremental
improvement drops below a preset threshold.

Initial clusters of points are almost completely determined by the
problem boundaries, making ImFil relatively insensitive to the initial
and allowing it to easily escape from local minima.
Conversely, this means that if the initial point is known to be of high
quality, ImFil must be provided with tight bounds around this point, or it
will unnecessarily evaluate points in regions that do not contain the global
minimum.

As a practical matter, for the noisy objective functions we studied, we find
that the total number of evaluations is driven almost completely by the
requested step sizes between successive clusters, rather than finding
convergence explicitly.

For {\sc scikit-quant} we have rewritten the original ImFil MATLAB
implementation into Python.

\subsection{SnobFit}
\label{ssec:SnobFit}

Stable Noisy Optimization by Branch and FIT (SnobFit)~\cite{snobfit} is an
optimizer developed specifically for optimization problems with noisy and
expensive to compute objective functions.
SnobFit iteratively selects a set of new evaluation points such that a balance
between global and local search is achieved, and thus the algorithm can escape
from local optima.
Each call to SnobFit requires the input of a set of evaluation points and
their corresponding function values and SnobFit returns a new set of points to
be evaluated, which is used as input for the next, recursive, call of SnobFit.
It is thus called several times in a single optimization step.
The initial set of points is provided by the user and should contain as many
expertly chosen points as possible (if too few are given, the choice is a
uniformly random set of points, and thus providing good bounds becomes important).
In addition to these points, the user can also specify the uncertainties
associated with each function value.
We have not exploited this feature in our test cases, because although we know
the actual noise values from the simulation, properly estimating whole-circuit
systematic errors from real hardware is an open problem.

As the name implies, SnobFit uses a branching algorithm that recursively
subdivides the search space into smaller subregions from which evaluation
points are chosen.
In order to search locally, SnobFit builds a local quadratic model around the
current best point and minimizes it to select one new evaluation point.
Other local search points are chosen as approximate minimizers within a trust
region defined by safeguarded nearest neighbors.\cite{snobfit}
Finally, SnobFit also generates points in unexplored regions of the parameter
space and this represents the more global search aspect.

For {\sc scikit-quant} we have rewritten the original SnobFit MATLAB
implementation into Python.

\subsection{BOBYQA}
\label{ssec:BOBYQA}

BOBYQA (Bound Optimization BY Quadratic Approximation)~\cite{bobyqa} has been
designed to minimize bound constrained black-box optimization problems.
BOBYQA employs a trust region method and builds a quadratic approximation in
each iteration that is based on a set of automatically chosen and adjusted
interpolation points.
New sample points are iteratively created by either a ``trust region'' or an
``alternative iterations'' step.
In both methods, a vector (step) is chosen and added to the current iterate to
obtain the new point.
In the trust region step, the vector is determined such that it minimizes the
quadratic model around the current iterate and lies within the trust region.
It is also ensured that the new point (the sum of the vector and the current
iterate) lies within the parameter upper and lower bounds.
BOBYQA uses the alternative iteration step whenever the norm of the vector is
too small, and would therefore reduce the accuracy of the quadratic model.
In that case, the vector is chosen such that good linear independence of the
interpolation points is obtained.
The current best point is updated with the new point if the new function value
is better than the current best function value.
Note that there are some restrictions for the choice of  the initial point due
to the requirements for constructing the quadratic model.
BOBYQA may thus adjust the initial automatically if needed.

Although it is not intuitively obvious that BOBYQA would work well on noisy
problems, we find that it performs well in practice if the initial parameters
are quite close to optimal and the minimum and maximum sizes of the trust
region are properly set.
This is rather straightforward to do for the specific case of VQE, where a
good initial guess can be obtained relatively cheaply from classical simulation.
For Hubbard model problems, which have many (shallow) local minima, BOBYQA
does not perform nearly as well.

In {\sc skikit-quant}, we use the existing PyBobyqa implementation~\cite{pybobyqa1, pybobyqa2}
directly.

\subsection{Validation and Tuning}
\label{sec:validation}

We have validated the {\sc scikit-quant} (re-)implementations for correctness
and performance using a suite of unit tests.
The defaults for each optimizer are chosen to work best for the type of problems
considered, and deviate from the defaults.

\comment{We have validated the {\sc scikit-quant} implementations for correctness and
performance using a suite of unit tests.
For ImFil and SnobFit, which have been ported from MATLAB, we have thoroughly
tested correctness, using their original tests as well as our own.
For NOMAD and PyBobyqa we invoke the original implementations, limiting the
need for testing beyond the application programming interface.
All tests have been included in the {\sc scikit-quant} repository.}

We have chosen defaults for each optimizer that should work best for the
type of problems considered.
Several of these choices deviate from the original defaults, and among others
involved an increase in the number of samples per iteration (PyBobyqa and
NOMAD in particular benefit here) and/or a tightening of the convergence
criteria (important for SnobFit).
This trades wall clock performance with science performance.
For ImFil, a reduction in the smallest step scales was needed, without which
chemical accuracy could not be achieved.
We balanced this cost with a reduction in the allowed number of internal
iterations in the interpolation on a stencil.

As a practical matter, choosing good hyperparameters is extremely important,
but too often, domain scientists tend to judge optimizers based on trial runs
on their problem at hand, rather than first studying their problem's
mathematical properties and only then searching for an optimizer to match,
with different tuning as needed.
This (faulty) approach may well cause them to miss out on the best choice.
Good, domain-specific, defaults should ameliorate this practical issue
somewhat.

\section{Hybrid Quantum-Classical Algorithms}
\label{sec:hqca}

\begin{figure}[ht]
\begin{minipage}{\columnwidth}
\centering
\includegraphics[width=0.9\columnwidth]{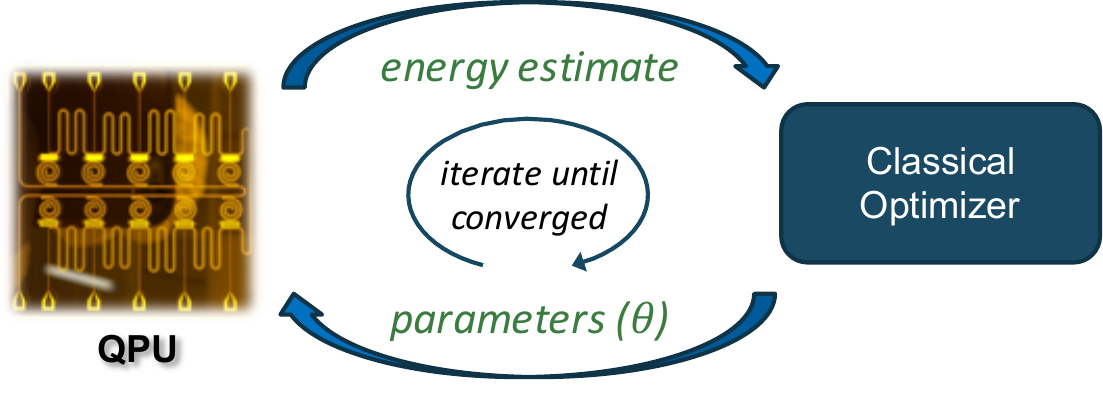}
\caption{\it \small VQE algorithm schematic.
The goal of the algorithm is to find
$\mathrm{E_0(\theta) = min_{\theta}(\langle\psi(\theta)|\mathcal{H}|\psi(\theta)\rangle/\langle\psi(\theta)\psi(\theta))}$
with the classical optimizer changing the input by varying optimization
parameters $\theta$ and the quantum chip calculating the expectation value
of $\mathcal{H}$.}
\label{fig:vqe}
\end{minipage}
\end{figure}

The hybrid quantum-classical algorithms we consider iteratively alternate
between a classical numerical optimizer and a quantum algorithm that
evaluates some objective to be minimized.
The classical optimizer varies a set of parameters that determine the input
state for the quantum processor to prepare.
The quantum side then executes an algorithm resulting in measurement and
some output distribution of probabilities.
This distribution is mapped into an objective function value that the
classical optimizer can handle, such as a single floating point number,
e.g.\ one representing the expected energy of a physical system
(see Figure~\ref{fig:vqe}).

In the VQE approach for solving chemistry problems, the objective function
calculates the expectation value of the Hamiltonian $\mathcal{H}$ associated
with a configuration of the simulated physical system.
Without noise, the optimization surface is expected to be smooth and convex
around the global minimum.
Bounds and constraints to help the optimizer and analysis are often
straightforward to obtain from physical laws, e.g.\ there should be no loss
of particles.

In Quantum Approximate Optimization Algorithms, the state is prepared
by a {\it p}-level circuit specified by {\it 2p} variational parameters.
Even at the lowest circuit depth (p=1), QAOA has non-trivial provable
performance guarantees.
Initial examples are from the domain of graph optimization problems,
e.g.\ MaxCut.
The optimization surfaces generated by QAOA problems can be arbitrarily
complex and bounds and constraints are harder to define as they need not be
physical.

Since we fully sweep noise levels up to the breaking points of the
optimizers, we believe that our findings are also applicable to any other
hybrid algorithms, including to the higher complexity in QAOA algorithms.
For more details, see Section~\ref{sec:related}.

\comment{Because of these last differences, understanding the impact of noise on the
behavior of hybrid algorithms is more straightforward for VQE and we will
concentrate our study on its behavior.
However, since we do not restrict the study to realistic noise levels only,
but push the optimizers to their breaking point, we believe that our
findings are directly applicable to the higher complexity in QAOA algorithms
as well.
For more details, see Section~\ref{sec:related}.}

\subsection{Role of the Ansatz in VQE}

The classical optimizer is not free to choose input states for VQE, but
constrained by a parametrized {\em ansatz}, which describes the range of
valid physical systems and thus determines the optimization surface.
A good ansatz provides a balance between a simple representation, efficient
use of available native hardware gates, and sufficient sensitivity with the
input parameters.
An effective ansatz can greatly reduce circuit depth, search space, and the
required number of iterations.

Ansatz design is still an art that requires detailed insights from the
domain science to uncover symmetries and to decide which simplifications
are acceptable.
However, our main interest is to push the optimizers.
Since a better ansatz will simply allow the domain scientist to work on
larger, more complex, problems that equally push the optimizer harder, we
will restrict ourselves to the commonly used, and practical, {\em unitary
coupled cluster} ansatz (UCC ansatz) for all studies.
For physical systems, the UCC ansatz can be thought of as describing the
movements of individual particles (linear terms) and those of interacting
(e.g.\ through electric charge) pairs of particles (quadratic terms).
It is simple to map and, because particles such as electrons are
indistinguishable, easy to find symmetries in to reduce the number of
parameters needed.

The choice of ansatz also affects the number of qubits used.
UCC has a natural, but somewhat inefficient, mapping.
More compact representations exist, but require changing the ansatz {\em and}
the operators, which can actually make the problem harder to solve as these
more complex operators are likely to require more native gates to implement.
Actual published results~\cite{Li_2017,Temme_2017,Dumitrescu_2018} comprise
only two and four qubit experiments with two parameters.
In our studies we have used 4 and 8 qubit problems, with the number of
parameters ranging from 2 to 14.

\comment{Besides the number of parameters, the choice of ansatz also affects the
number of qubits used.
For example, the UCC ansatz provides for simple physical interpretations,
such as `1' meaning that a site or orbital is occupied by an electron, and
`0' meaning that it is unoccupied.
Add a second qubit for spin up and down, and two qubits can fully describe a
site or orbital.\footnote{It is still completely up to the domain scientist to
determine which and thus how many sites are relevant for the problem they are
trying to solve, which is the most important driver of the number of qubits
needed.}
However, there is a clear inefficiency here: it is unnecessary to describe
the spin of an unoccupied site.
But changing to a more compact representation requires changing the ansatz
and the operators, which can actually make the problem harder to solve.
Published results~\cite{Li_2017,Temme_2017,Dumitrescu_2018} comprise only two
and four qubit experiments with two parameters.
In our studies we have used 4 and 8 qubit problems, with the number of
parameters ranging from 2 to 14.}

\subsection{VQE Quantum Processor Step}

The quantum circuit consists of a state preparation and a time evolution
step.\cite{RevModPhys.92.015003}
By representing the Hamiltonian as a product of Pauli operators, it can be
split across its components to be evaluated independently and combined on
the classical side.
The chip readout is a probability distribution of bit strings that represent
the contributions of each component under the chosen encoding.
Although the variance due to the probabilistic nature of the measurement can
be estimated\cite{ibmnature}, the effect of on-chip noise depends on its
effect on the projective measurement and hence on the output state and the
encoding.
Furthermore, typical post-processing steps such as e.g.\ Bayesian unfolding
to ensure that the measured probability distributions are physical (and sum
to 1), mean that the final mapping to a single number (the energy to be
minimized) is no longer a linear operation in the presence of on-chip noise.
It is thus not possible to make any {\em general} inference about the
uncertainty distribution of the estimated energy from the expected errors in
the probability distribution, but only about {\em specific} problem instances.

\section{Impact of Noise}
\label{sec:noise_impact}

VQE is considered to have some robustness against noise due to its iterative
nature and hence is expected to be well suited for upcoming NISQ devices.
Nevertheless, the need for studying the dynamics of the full hybrid VQE
algorithm has been identified early on~\cite{McClean2015} as a prerequisite
for running it successfully on NISQ hardware.

\subsection{Accounting for Noise Sources}

There are a range of ways that noise affects the final result, but the
exact mechanisms are an area of open research and there are no accurate
predictive models available yet.
Our main concerns, however, are about overall magnitude of noise and the
effects on the shape of the optimization surface.
We thus cover the problem domain by varying the magnitude of the noise in
simulation by a wide range, and by studying different problems with
different optimization surfaces.
Actual noise impact for any given hardware instance is likely captured within
our parameter sweep.
Our intent is to arrive at a {\em map and guidance} for actual experiments.
The goal is explicitly not to find and describe \textit{the} single way, if
any such exists, of how VQE behaves with a given noise model, nor to find the
one optimizer to use for all VQE problems.
It is well known in the applied math community that there is no
such thing as a ``free lunch,'' meaning that each optimizer has specific
strengths, none are best in all instances, and each problem needs to be
individually matched to the appropriate optimizer(s).

We take an empirical approach, injecting noise as Gaussian-distributed
over-/under-rotations with an added orthogonal component onto the circuit
gates.
This ensures realistic properties: noise increases with circuit depth and
complexity, and two-qubit gates have larger contributions than one-qubit
gates.

We do not add coherent or correlated noise sources, for the reasons
explained in the next section, but also because orthogonal error mitigation
techniques such as Randomized Compiling~\cite{PhysRevA.94.052325} have been
shown to alleviate coherent errors by making them stochastic.
We also do not add measurement errors, because shot noise is expected to be
unbiased and can be reduced by increasing the number of measurements.

\comment{The measurement result is a probability distribution of bit strings, and any
type of stochastic noise behaves on it in a similar way: they proportionally
redistribute relative counts, and have the same equilibrium in the limit
(a uniform distribution).
Coherent and correlated noise sources can, however, result in {\em any}
biased distribution, making their study meaningless, unless taken from the
behavior of actual hardware.
But doing so would limit their relevance to that specific hardware.
Further, as detailed below, VQE has more ``builtin'' robustness against
coherent than against stochastic noise.
Coherent noise can also be expected to more easily produce non-physical
outcomes (e.g.\ fewer or more particles in the final than in the input
states); those measurements can be filtered out and discarded.
Last but not least, orthogonal error mitigation techniques such as Randomized
Compiling~\cite{PhysRevA.94.052325} have been shown to alleviate coherent
errors by making them stochastic.}

We do not factor in an additional noise contribution from measurement errors:
shot noise is expected to be unbiased (i.e.\ it can be averaged out to zero
noise in the limit by taking a large number of measurements).
In other words, it affects the overall magnitude of stochastic noise sources,
which we already sweep, not what we most care about: the shape changes in the
optimization surface.

\subsection{Interplay with Minimizer}

Some general observations can be made about the different impacts of coherent
and stochastic errors, and why the distinction matters on hybrid
quantum-classical algorithms that involve a classical optimizer, such as VQE.

Quantum computing is very sensitive to noise, because a noisy execution is just
as valid as a noise-free one: without error correction codes, there is no
distinguishing between valid and erroneous states.
Therefore, if a circuit is intended to simulate the evolution of some
Hamiltonian $\mathcal{H}$, then a single noisy run can be seen as the
evolution of some other Hamiltonian $\mathcal{H}'$.
As long as the noise level is ``small enough,'' the eigenstates of
$\mathcal{H}$ and $\mathcal{H}'$ will be close.

\begin{figure}[h]
\begin{tabular}{cc}
\begin{minipage}{0.46\columnwidth}
\includegraphics[width=\textwidth]{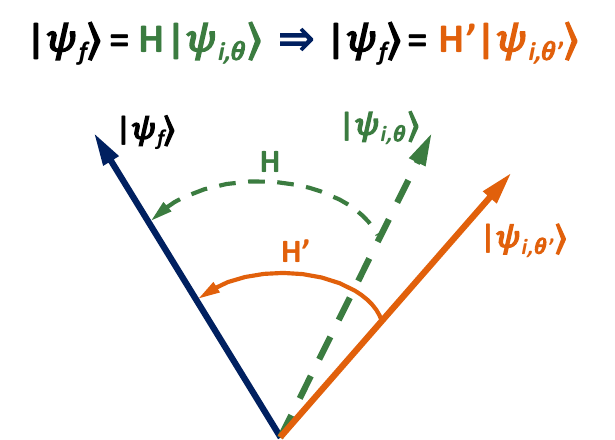}
\end{minipage}
  &
\begin{minipage}{0.46\columnwidth}
\centering
\includegraphics[width=\textwidth]{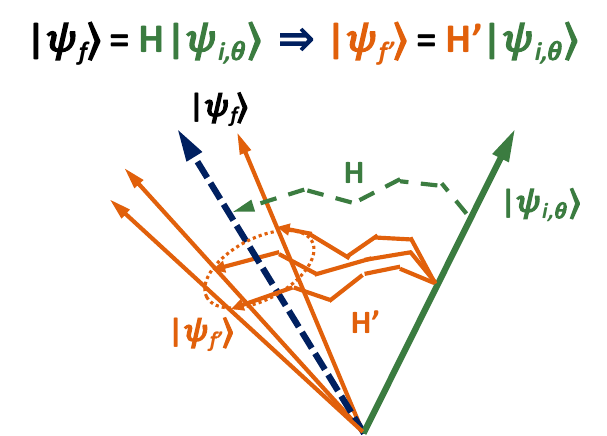}
\end{minipage}
 \\
\end{tabular}
\caption{\it \small
For coherent noise, the optimizer adjusts the input for its predictable
impact and still finds the global minimum (left).
Stochastic noise ``randomly walks'' away from the intended output state
(right), with the global minimum increasingly unlikely to be reachable.}
\label{fig:noise_impact}
\end{figure}

\textbf{The algorithm is somewhat robust to coherent errors.}
Systematic errors are predictable, allowing the optimizer to find the global
minimum by compensating in the choice of input state.
At that minimum, changes in the output state are zero by definition for small
deviations in the input state, and the eigenvalue is classically calculated
with the correct $\mathcal{H}$.
Thus, the  minimum energy will still be very close, but the optimal
parameters found will be systematically off, see
Figure~\ref{fig:noise_impact} (left).
Specific to VQE, the ansatz restricts the input states that can be chosen,
thus VQE is more easily impacted by coherent errors than hybrid algorithms in
general.

\textbf{The algorithm has challenges with stochastic noise.}
The picture changes significantly with stochastic noise: each execution of
the circuit is in effect a different $\mathcal{H}'$.
Once close to the global minimum, the minimizer will not be able to distinguish
the outputs of runs with different inputs, as the changes get washed out in
the noise (see e.g.\ Figure~\ref{fig:hubbardnoise}).

\begin{figure}[h]
\centering
\vspace{-0.15in}
\includegraphics[height=2.1in]{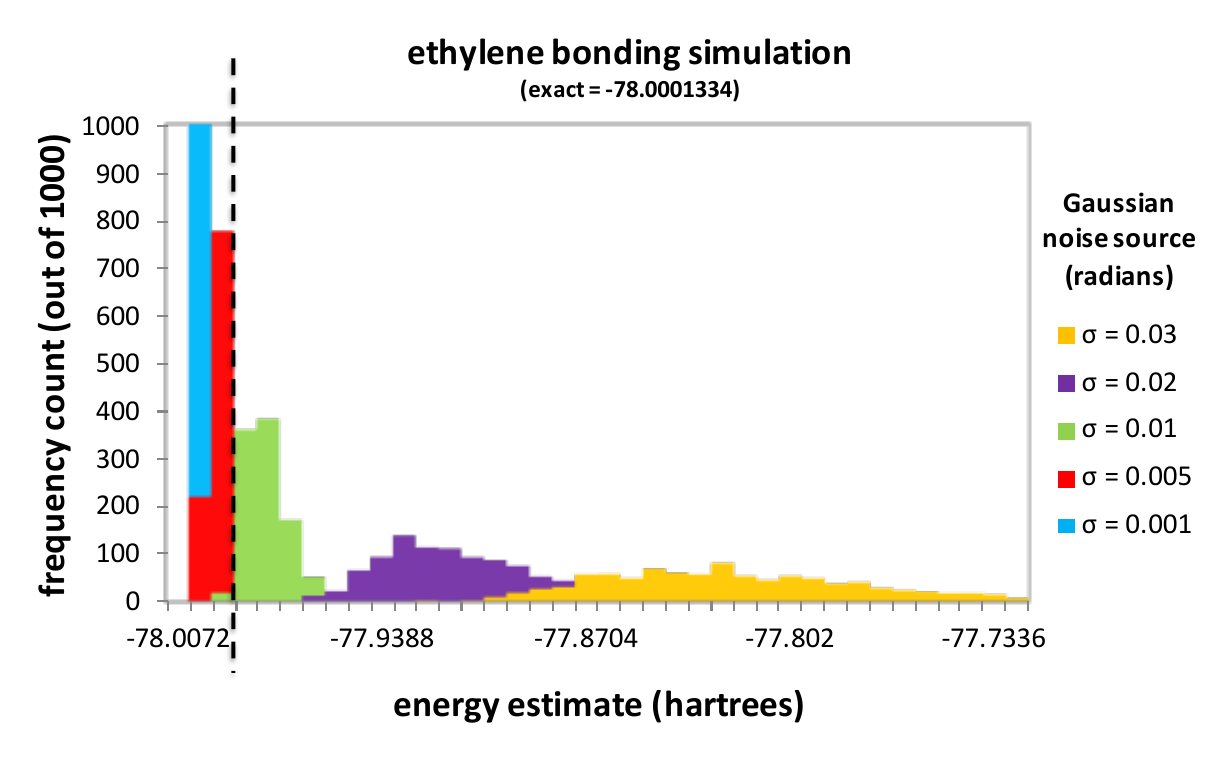}
\vspace{-0.1in}
\caption{\it \small Evaluation at the optimal parameters for ethylene bond
breaking simulation with increasing stochastic errors.
Distributions broaden, the average lifts, and eventually the true global
minimum is no longer reachable.
The dashed line shows the cut-off for chemical accuracy for useful
scientific results.}
\label{fig:objnoise}
\end{figure}

With sufficient symmetry in the optimization profile or a functional
description based on the domain science, the optimizer can still find the
correct optimal parameters by searching for a {\em robust} global minimum or
doing a local fit.
However, any execution at the optimal parameters will calculate an output
distribution that is some random walk away from the intended state, as the
errors (in particular those on the control qubit of CNOTs) do not commute
with the circuit as a whole, see Figure~\ref{fig:noise_impact} (right).
When calculating the energy objective from any of these noisy outputs that
are close to, but not at, the global minimum, the results will by definition
be higher than the ground state energy\footnote{Unless the output state no
longer represents the initial physical system, e.g.\ if electrons are lost
or spins have flipped in the simulation due to T1 noise.
In that case the energy found can end up below the ground energy of the
intended Hamiltonian.}
With increasing noise, the likelihood of the true global minimum energy
being returned by the objective function goes to zero, as shown in
Figure~\ref{fig:objnoise}.

\comment{
\begin{figure}
\tiny
\bgroup
\def\arraystretch{1.1}
\begin{tabular}{ c c | c | c | c  }
\multirow{12}{*}{
\begin{minipage}{0.35\columnwidth}
\centering
\includegraphics[width=\textwidth]{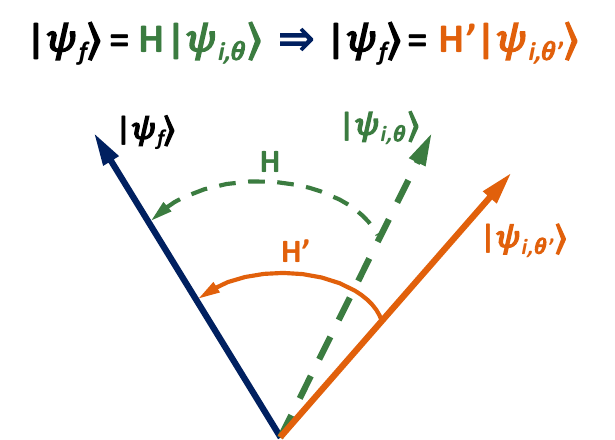}
\end{minipage}}
  & \multicolumn{4}{c}{4-qubit VQE, ethylene rotation simulation}        \\
  & \multicolumn{4}{c}{2 parameters; ``science cut-off" at -77.93318 ha} \\
  & {\bf $\mu_{noise}$} {\bf (rad.)} & {\bf $\theta_{0}$} & {\bf
                                                            $\theta_{1}$}
                                     & {\bf energy (ha)} \\[1ex]
 \cline{2-5}
  & 0.0001 & 9.46E-7 & 0.1778 & -77.93478 \\
  & 0.0005 & 1.47E-6 & 0.1778 & -77.93477 \\
  & 0.001  & 4.08E-6 & 0.1778 & -77.93476 \\
  & 0.005  & 8.81E-5 & 0.1779 & -77.93442 \\
  & 0.007  & 1.73E-4 & 0.1779 & -77.93389 \\
  & 0.01   & 4.55E-4 & 0.1780 & -77.93338 \\
 \cline{2-5}
 \cline{2-5}
  & 0.02   & 1.47E-3 & 0.1785 & -77.92924 \\
  & 0.03   & 3.48E-2 & 0.1793 & -77.92247 \\
\end{tabular}
\egroup
\caption{\it \small Effect of systematic errors.
The end result is the same for the same input parameters, thus if the error
is small, the global minimum can still be reached by the minimizer adjusting
the input to counter the systematic error.
The end results are incorrect parameters, but a correct minimum value.}
\label{fig:systematic_err1}
\end{figure}
}

\comment{
\begin{figure}
\tiny
\begin{tabular}{ l l }
\begin{minipage}{0.35\columnwidth}
\centering
\includegraphics[width=\textwidth]{figures/effect_stochastic}
\end{minipage}
  &
\begin{minipage}{0.6\columnwidth}
\centering
\includegraphics[width=\textwidth]{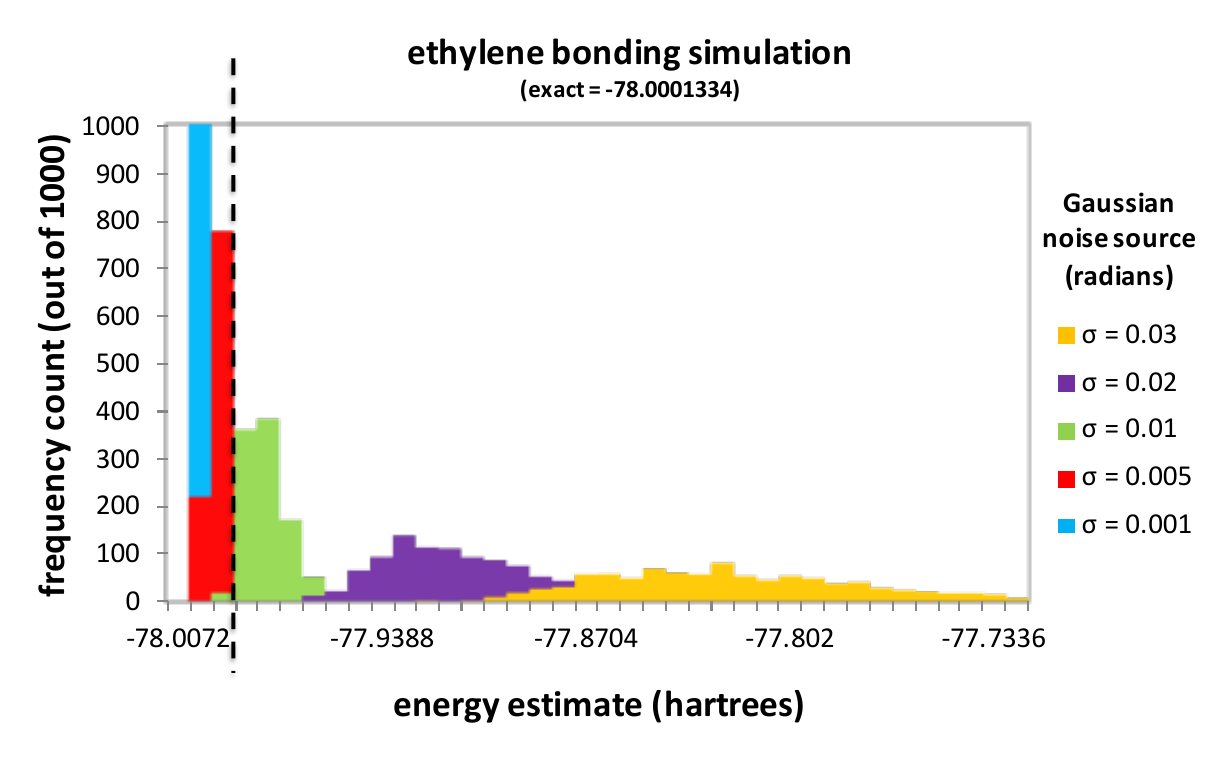}
\end{minipage}
 \\
\end{tabular}
\caption{\it \small Effect of stochastic errors.
The end result will vary even when given the same input parameters, and the
average result is a ``random walk" away of the error-free result.
Energy estimates broaden in distribution and the average lifts.
Eventually, with increasing noise, the true global minimum is never
returned.}
\label{fig:stochastic_err1}
\end{figure}
}

\section{Results}
\label{sec:results}

\begin{figure}[h]
\centering
\includegraphics[height=1.5in]{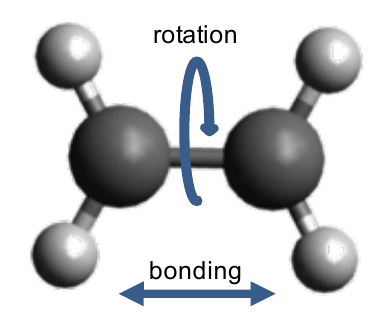}
\caption{\it \small Illustration of the ethylene rotation/torsion and bond
stretching/breaking simulations.}
\label{fig:ethybench}
\end{figure}

As study cases, we used the C-C axis rotation and bond stretching and
breaking of the ethylene ($C_{2}H_{6}$) molecule (see
Figure~\ref{fig:ethybench}), representing two different chemical
transformation processes.
In the rotation and bonding processes, the character of the wave function
changes drastically.
For example, in the C-C axis rotation $\Pi-\Pi$ bonds are broken/formed.

\begin{figure}[h]
\centering
\includegraphics[height=1.5in]{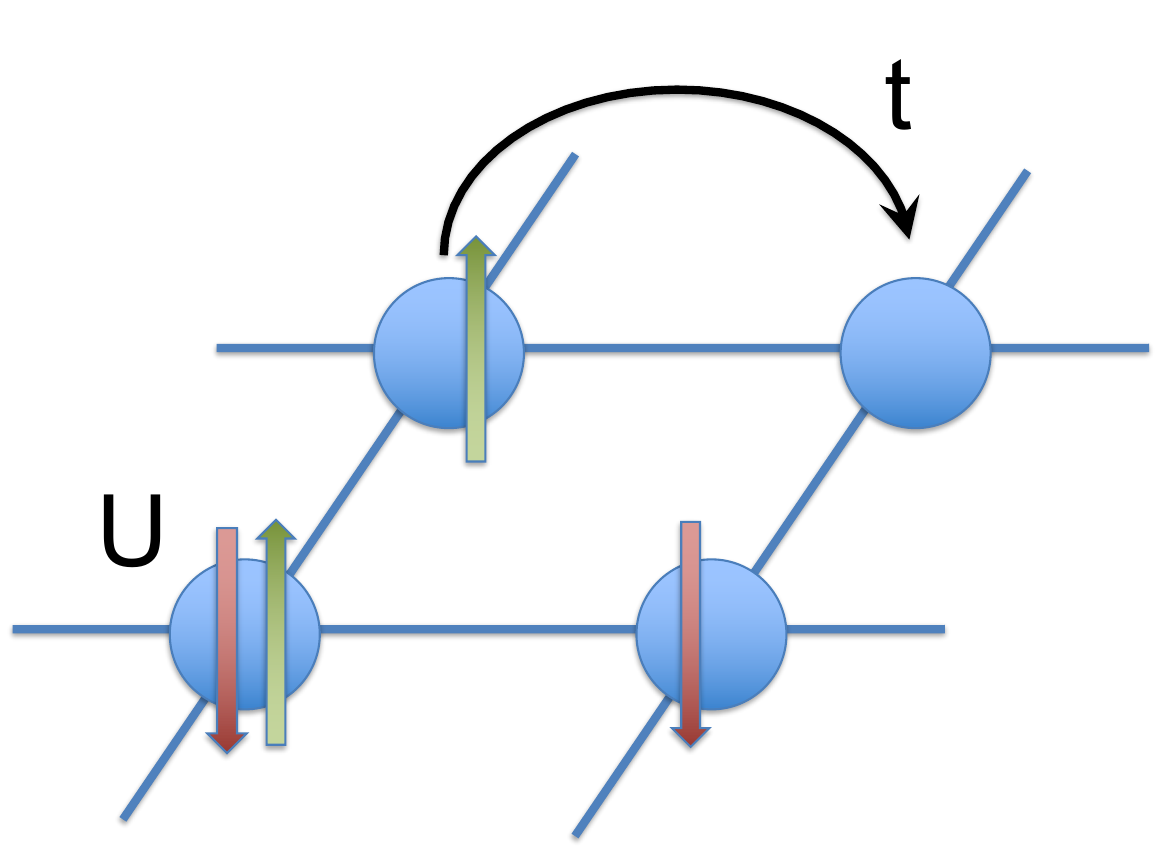}
\caption{\it \small Illustration of the 4-site Hubbard model simulation for
electrons with spins, using hopping term $t$ and Coulomb term $U$.}
\label{fig:hubbbench}
\end{figure}

We also used a Hubbard simulation of 4 sites, occupied with either 4 or 2
electrons (see Figure~\ref{fig:hubbbench}).
In the Hubbard simulations, we use a hopping term of $1.0$, a Coulomb term of
$2.0$, and in the 4 electron case add a chemical potential of $0.25$.
The electrons always have spins.
In all cases, OpenFermion~\cite{openfermion} is used to generate the
circuits.

With a Unitary Coupled Cluster ansatz (see Section~\ref{sec:hqca}), the
minimal representation to describe the rotation simulation consists of 4
qubits (i.e.\ 4 orbitals) and 2 terms in the wave function expansion to
optimize.
Similarly, the bond breaking process requires 8 qubits and uses 14 parameters
in the expansion.
The 4 site Hubbard model requires 8 qubits and 9 parameters for a 2 electron
occupancy; and 8 qubits with 14 parameters when simulating 4 electrons.

\subsection{Experimental Setup}

{\bf Noise Injection:}
We extended the ProjectQ~\cite{projq} quantum simulation infrastructure with
noise injection capabilities.
For each gate in the circuit circuit ($R_{X}(\theta)$, $R_{Y}(\theta)$, $H$,
$CNOT$\footnote{We add no noise to $R_{Z}(\theta)$: these are purely
mathematical, thus noise-free.}), we add an operator in the form of rotations
whose angles are independently sampled from a probability distribution.
For each scenario we perform sweeps with increasing noise strength until it
breaks the minimizers.
In the rest of this paper, numerical values for noise magnitude refer to the
standard deviation ($\sigma$) of the Gaussian noise probability distribution.

{\bf Methodology:}
In each study, the minimizer is given an appropriate budget (maximum number
of invocations of the objective function) and convergence criteria are
adjusted in favor of using up the budget.
The minimizers are run until any convergence criteria are met or the budget
is used up.
We repeat the full algorithm several times and report the average and
overall minimum across all runs, as well as the averaged result when running
the simulation at the optimal parameters found.
The results are compared to the results of classical ab-initio calculations.

{\bf Optimizer Baseline:}
The optimizers included in {\sc scikit-quant} have been described in
Section~\ref{sec:scq}.
Each optimizer has been individually tuned with good settings for the type
of problems generated by our VQE test circuits (see
Section~\ref{sec:validation}).
As baseline comparisons, we choose BFGS and Cobyla, both from
SciPy~\cite{scipy_web}, because they are well known and widely used, as
explained in Section~\ref{sec:baseline}.

{\bf Hardware:}
The simulations were small enough, memory-wise, to run on a standard server.
We note that for this study simulating the quantum circuit constitutes the
main bottleneck; optimizers can run well and handle a large number of
parameters when using just a single compute node.

\subsection{Optimization Solution Quality}
\label{sec:optreq}

\begin{figure*}
\begin{tabular}{cc}
\begin{minipage}{\columnwidth}
\centering
\includegraphics[width=1.0\columnwidth,height=1.98in]{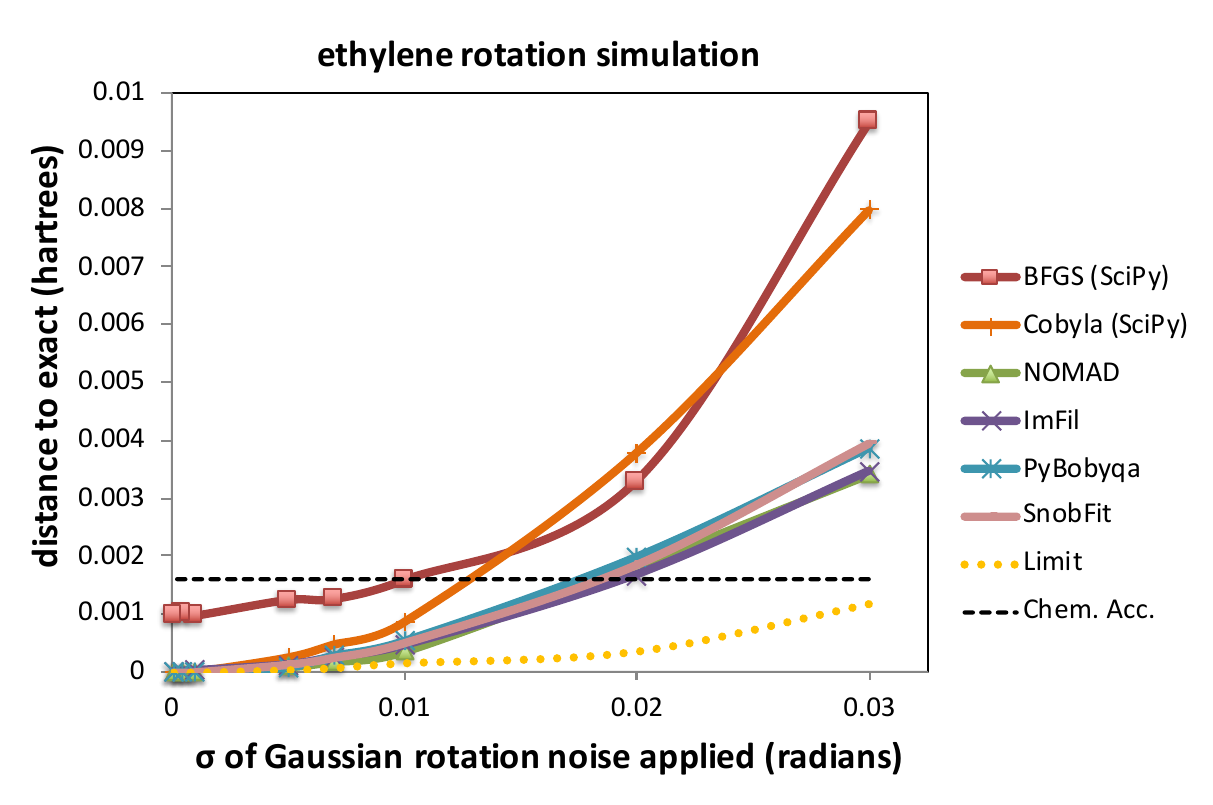}
\end{minipage} &
\begin{minipage}{\columnwidth}
\centering
\includegraphics[width=1.0\columnwidth,height=1.98in]{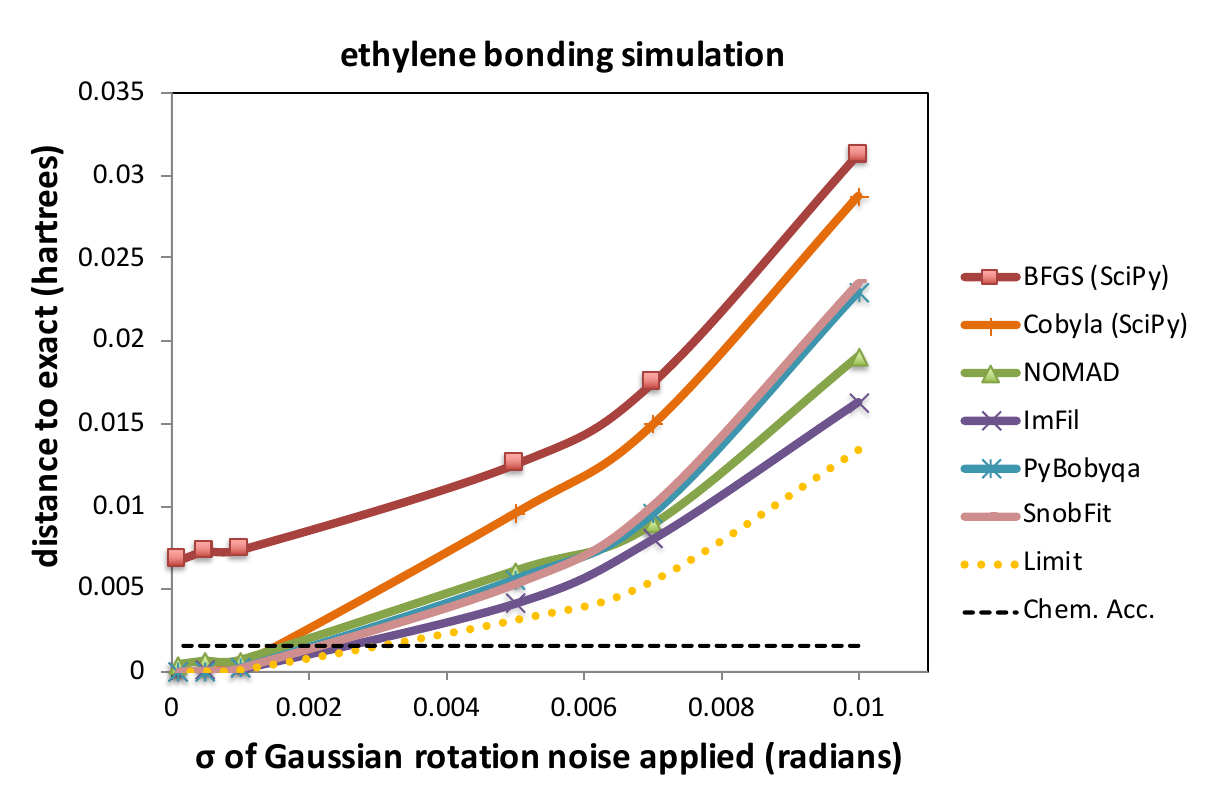}
\end{minipage} \\
\end{tabular}
\vspace{-0.10in}
\caption{\label{fig:ethy_results}\it \small Average calculated energy
of the full VQE algorithm for the ethylene rotation (left) and bond breaking
simulation (right).
Lowest noise level is $10^{-4}$.
The cut-off for chemical accuracy is shown by the straight, dashed, black
line.
With increasing noise, the result from the objective function is increasingly
moved away from the global minimum.
The lowest value that the objective function could return at a given noise
level is estimated by the dashed yellow line.}
\end{figure*}

\begin{figure*}
\begin{tabular}{cc}
\begin{minipage}{\columnwidth}
\centering
\includegraphics[width=1.0\columnwidth,height=1.98in]{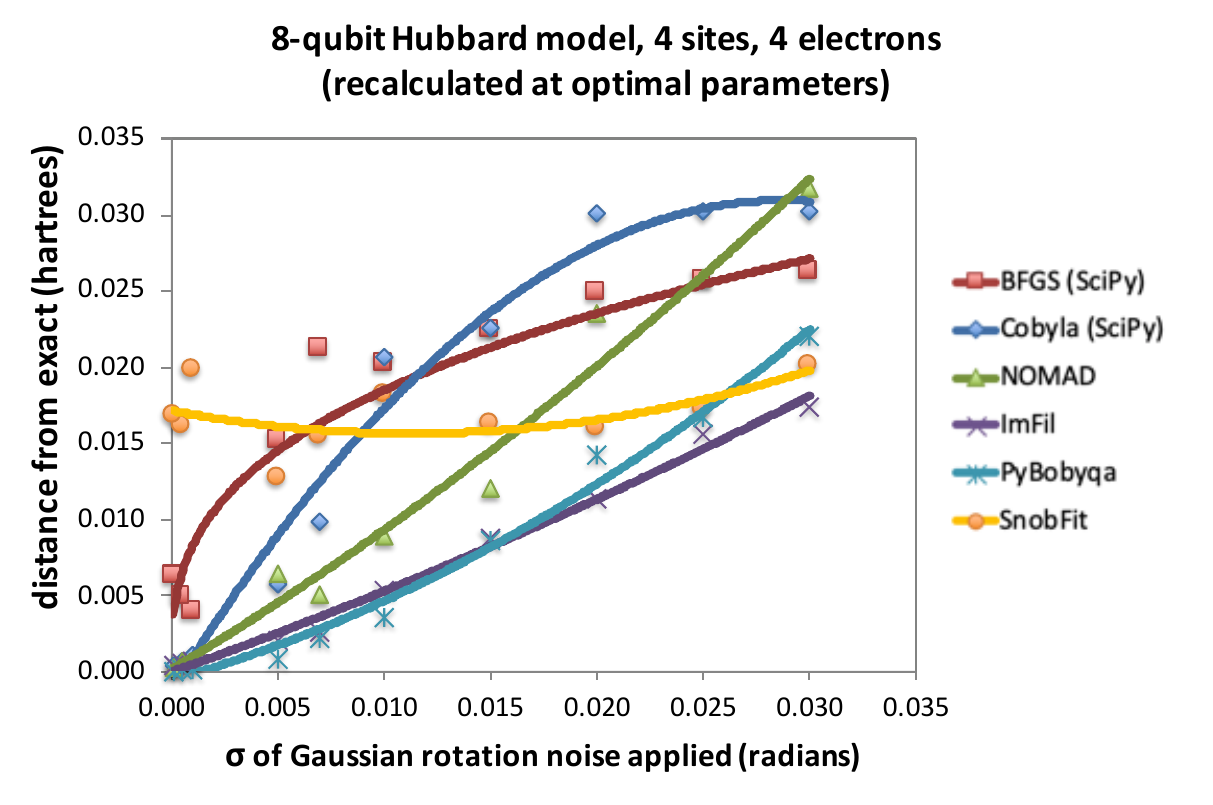}
\end{minipage} &
\begin{minipage}{\columnwidth}
\centering
\includegraphics[width=1.0\columnwidth,height=1.98in]{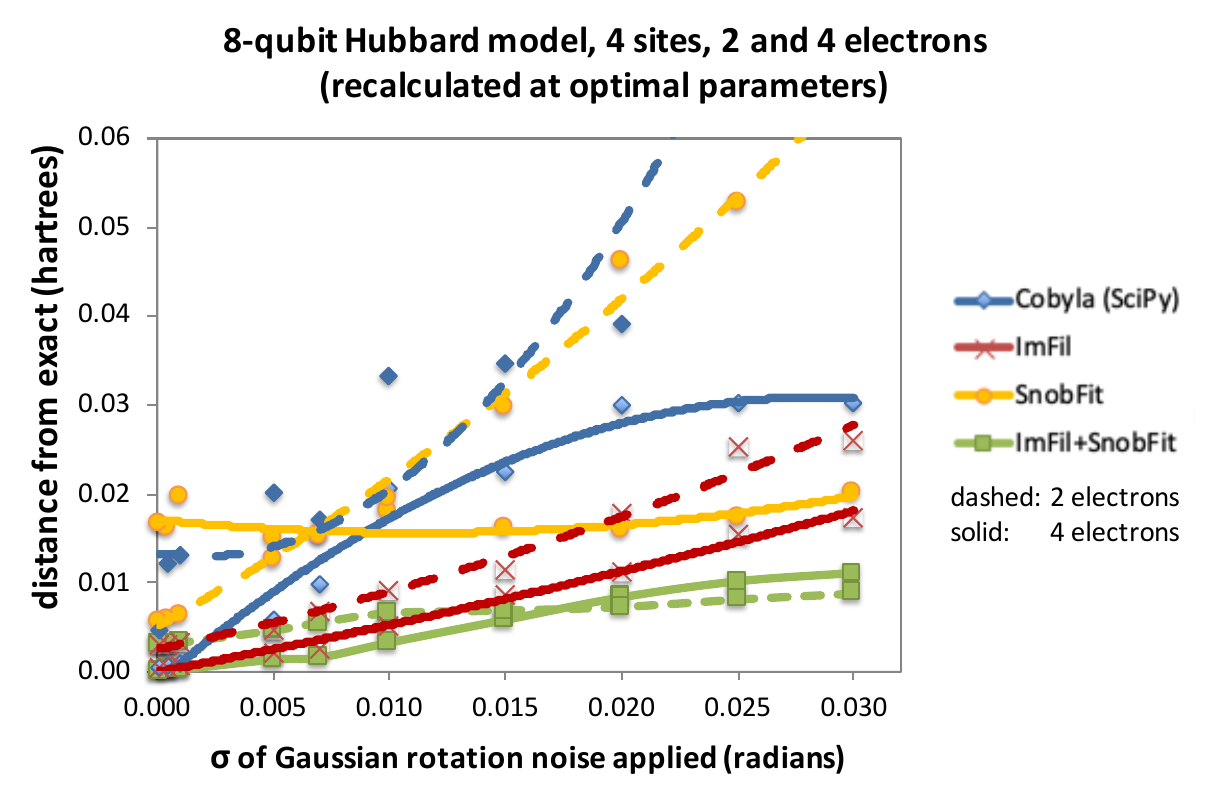}
\end{minipage}\\
\end{tabular}
\vspace{-0.10in}
\caption{\label{fig:hubb_results}\it \small Full VQE results of a 4
  site Hubbard model simulation with 4 electrons for individual optimizers
  (left); and results when combining ImFil and SnobFit for 2 (dashed lines)
  and 4 (solid) electrons (right).
  The ground energy is recalculated at the optimal parameters found using a
  noise-free simulation.
  The plotted results are therefore quadratic regression lines fitted to the
  data, which showed great variability.}
\end{figure*}

\begin{figure*}
\begin{tabular}{cc}
\begin{minipage}{\columnwidth}
\centering
\includegraphics[width=1.0\columnwidth,height=1.98in]{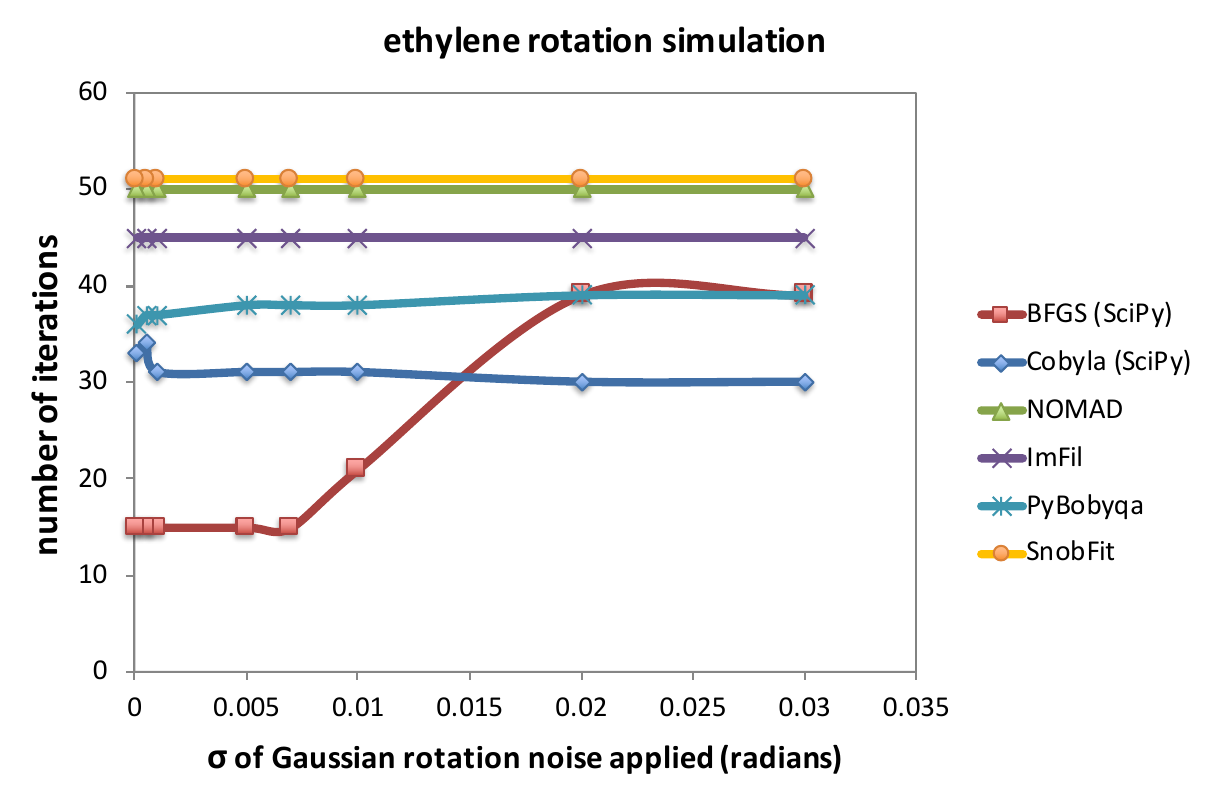}
\end{minipage} &
\begin{minipage}{\columnwidth}
\centering
\includegraphics[width=1.0\columnwidth,height=1.98in]{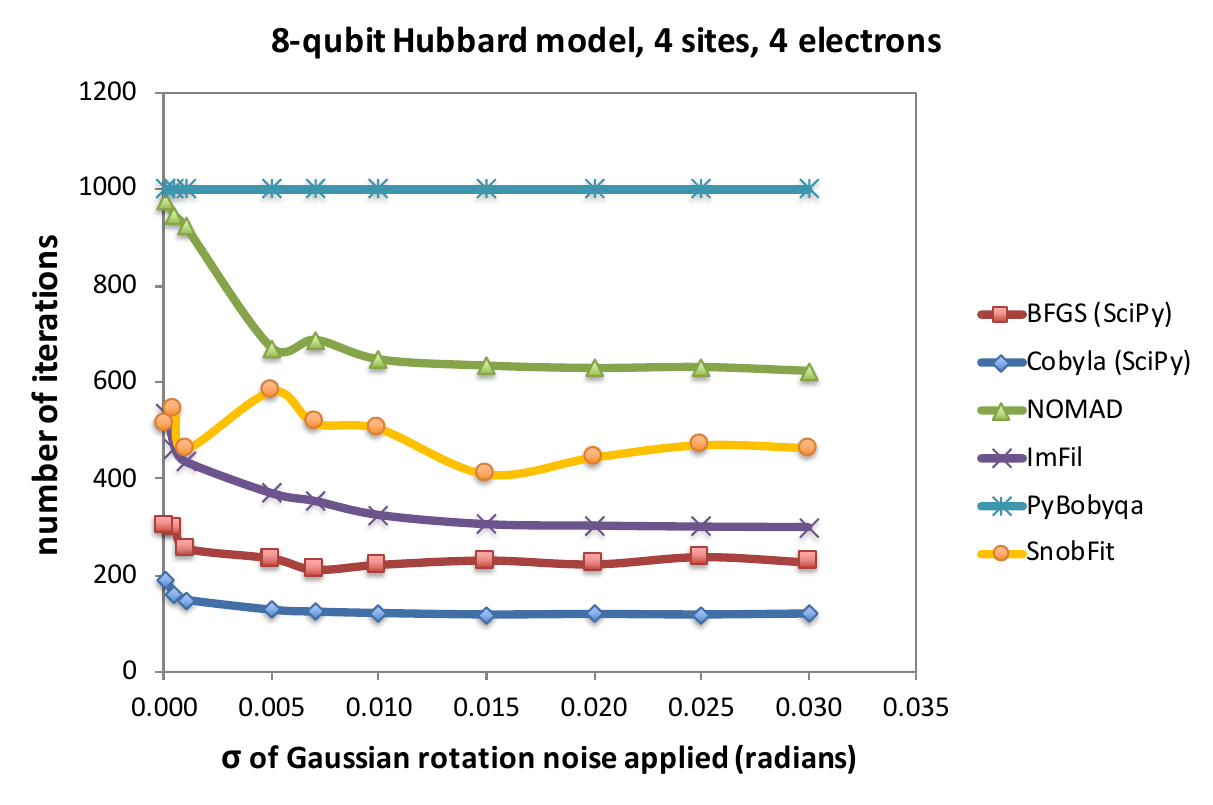}
\end{minipage} \\
\end{tabular}
\vspace{-0.10in}
\caption{\label{fig:iterations}\it \small Number of objective
  function calls used for each of the optimizers as a function of the
  noise level.
  Ethylene rotation simulation (left; budget of 50) and Hubbard Hubbard
  model simulation with 4 electrons (right; budget of 1000).}
\vspace{-0.12in}
\end{figure*}

\begin{figure*}
\begin{tabular}{cccc}
\begin{minipage}{0.53\columnwidth}
\hspace{-0.2in}
\includegraphics[width=\columnwidth]{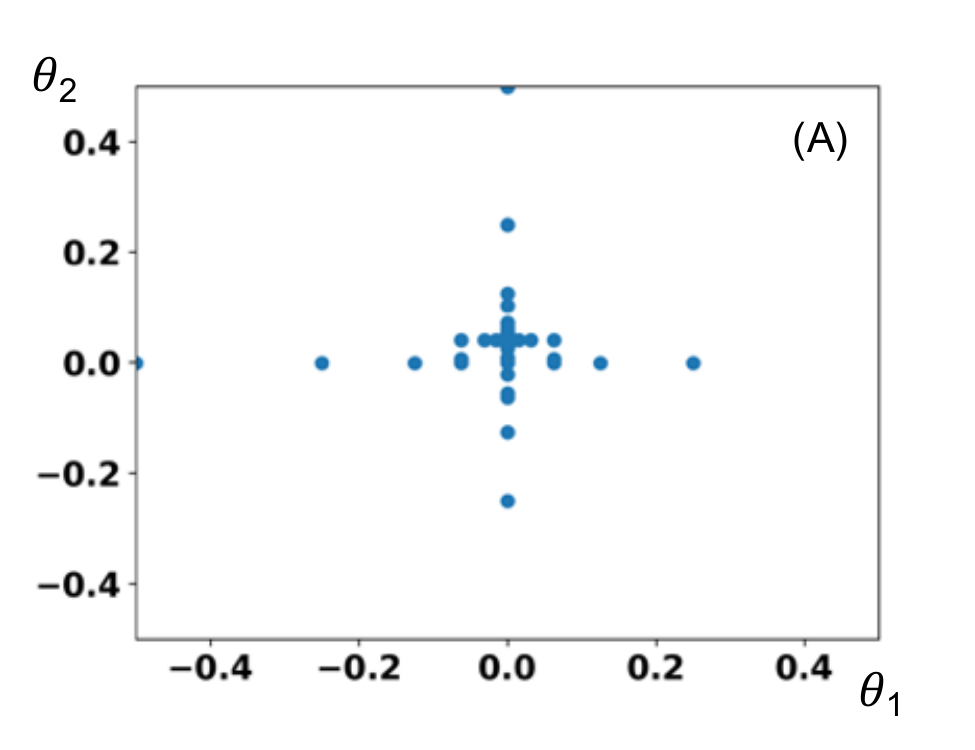}
\end{minipage} &
\begin{minipage}{0.53\columnwidth}
\hspace{-0.3in}
\includegraphics[width=\columnwidth]{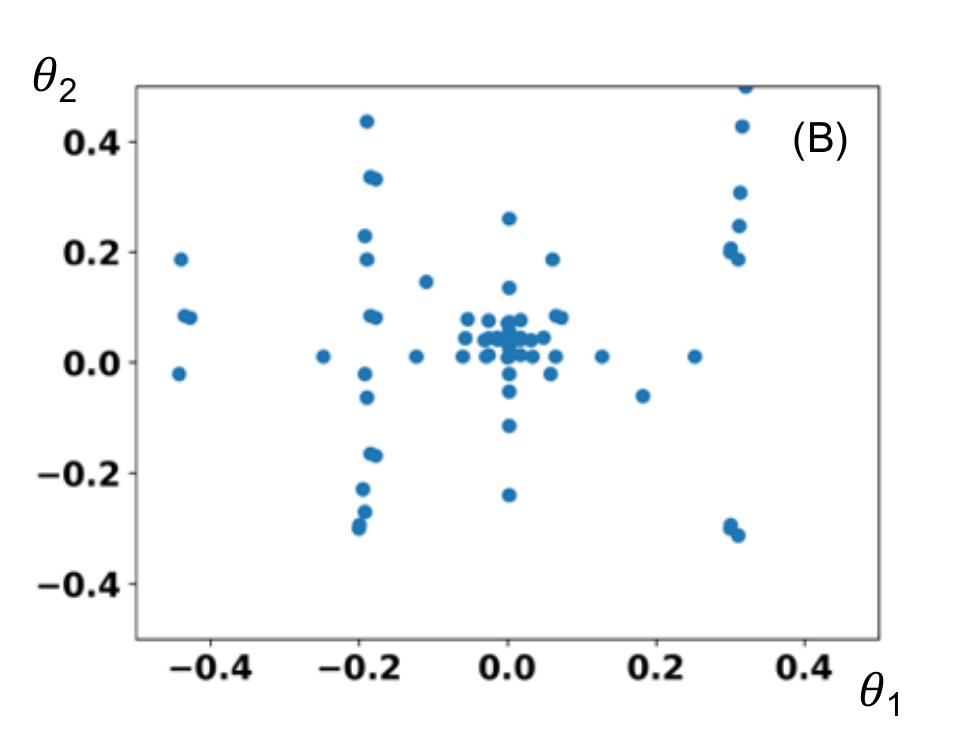}
\end{minipage} &
\begin{minipage}{0.53\columnwidth}
\hspace{-0.5in}
\includegraphics[width=\columnwidth]{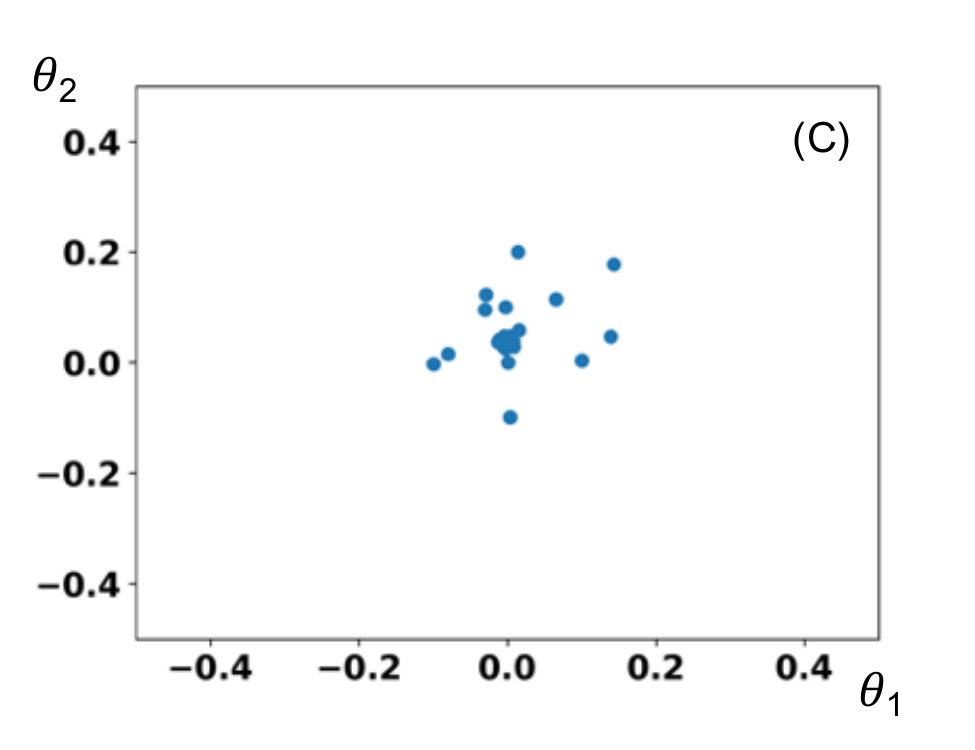}
\end{minipage} &
\begin{minipage}{0.53\columnwidth}
\hspace{-0.7in}
\includegraphics[width=\columnwidth]{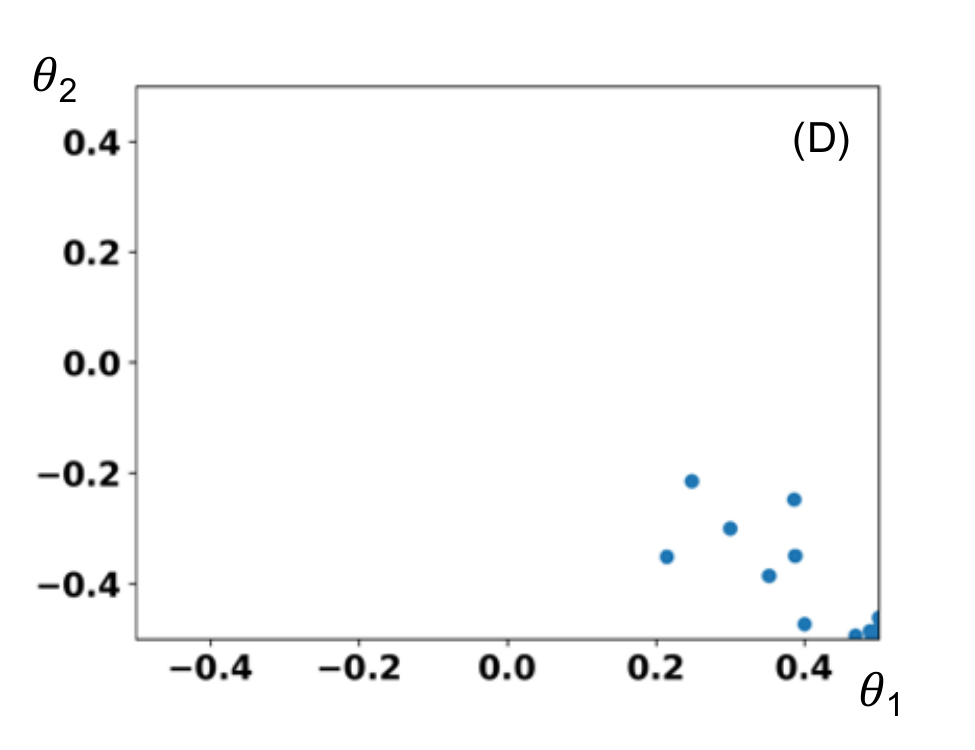}
\end{minipage} \\
\end{tabular}
\vspace{-0.1in}
\caption{\label{fig:search}\it \small Effect of the quality of the
initial on the overall solution for the ethylene rotation simulation.
Shown are parameters chosen by the optimizer to evaluate: ImFil with good (A)
and with bad initial (B); PyBobyqa with good (C) and bad initial (D).
The good initial is at $(0.1, 0.1)$, the bad at $(0.3, -0.3)$, and the global
optimum is at $(0.00012, 0.04)$.
ImFil requires more evaluations for its global search, but does not risk
getting trapped in a local minimum $(0.5, -0.5)$ like PyBobyqa.}
\end{figure*}

There are two ways to evaluate the optimizers: 1) by the minimum energy they
actually find relative to what was possible given the response limitations of
the objective function; or 2) by the quality of the optimal parameters found,
evaluated by recalculating the expected energy from a noise-free simulation
run at those parameters.
Which quality measure is most relevant will depend on the application and
science goals at hand, so we provide examples of both.
E.g.\ in the case of chemistry studies, quantum subspace
expansion~\cite{McClean_2020} requires accurate parameters more than an
accurate value.

\textbf{Distance to minimum energy.}
Figure~\ref{fig:ethy_results} shows the average calculated energy of the full
VQE algorithm for the ethylene rotation (left) and bond breaking simulation
(right), for 100 runs at each noise level for the former and 10 each for the
latter.\footnote{The 8-qubit circuits took about two orders of magnitude more
time to run.}
The straight, dashed, black lines show the chemical accuracy ($0.00159$
hartrees): a solution closer to the exact value than this cut-off
(i.e.\ results below this line) are scientifically useful.
The dashed yellow lines show the lowest value the objective function returned
across all runs, i.e.\ the lowest value any of the minimizers could
theoretically have found.
Where this line is above the chemical accuracy, {\em the optimizer is not the
weak link of the algorithm}, the quantum processor is the limiting component.
The larger, deeper, 8-qubit circuit clearly suffers more from noise: even at
moderate levels, a chip with such gate noise would be the weak link in the
full algorithm.

Of the minimizers, BFGS can not find the global minimum even at low levels
of noise (lowest shown is $10^{-4}$), because it treats any gradients seen as
real, including fakes due to noise, and gets stuck.
It worked, however, fine on a noise-free run (not plotted).
Cobyla, the other baseline, performs quite well at low levels of noise, but
clearly underperforms as noise increases.
The optimizers designed to handle noise well outperform across the full
range, with some stratification only happening at the highest noise levels
and ImFil yielding the overall best results.
In the low noise regime, however, where all optimizers perform similarly,
other considerations, such as the total number of iterations, come into play
to determine which is ``best.''
Cobyla would then most likely be preferred
(see Section~\ref{sec:perf_considerations} for a detailed discussion).

\textbf{Parameter quality.}
Figure~\ref{fig:hubb_results} (left) shows the results for the full VQE
algorithm Hubbard model simulations, with the energy recalculated at the
optimal parameters using a noise-free run.
With the Hubbard model, the region of the optimization surface around the
global minimum is rather shallow (see also Figure~\ref{fig:hubbardnoise}),
which clearly stresses the optimizers a lot more.
The behavior of BFGS and Cobyla mimics the results from the ethylene
studies, but this time both NOMAD and especially SnobFit also underperform
or even fail.
A detailed analysis shows that this weakness is exposed by bounds that are
too large for either optimizer to handle: reducing the bounds greatly
improves their performance (whereas it does not for BFGS and Cobyla).

\subsection{Leveraging Domain Science Knowledge}

It is already apparent that different methods perform best for different
problems as optimization surfaces vary.
Furthermore, the quality of the solution may be improved by exploiting a
combination of domain science and optimizer knowledge.
The most obvious and realistically actionable are: 1) quality of
initial solution; and 2) good parameter bounds.

\textbf{Impact of initial solution quality.}
For chemical problems, a good initial can be obtained from approximate
classical calculations.
To understand its impact, we consider a comparison of ImFil and PyBobyqa
for the ethylene rotation simulation.

In Figure~\ref{fig:search} we plot the evaluation points chosen by each
optimizer: using a good initial at $(0.1, 0.1)$ and a bad one at
$(0.3, -0.3)$.
The  global optimum is at $(0.00012, 0.04)$.
Whether it receives a good (A) or bad (B) initial, ImFil will use the given
bounds to determine its first stencil, doing a mostly global search.
Although the initial drives the first few iterations, it quickly moves away
from the bad initial, to converge at the optimum.
PyBobyqa starts by considering only points within its trust region around
the initial point.
If the initial is close enough to make the global optimum fall within that
region, it will find it quickly (C).
However, if the initial is near a pronounced local minimum, in $(0.5, -0.5)$
in this case, it will get stuck (D), never finding the global minimum.

\comment{Overall, this analysis indicates that if good initials are available
with low computational overhead, they can improve both the quality and
speed to solution.}

{\it \bf Impact of bounds.}
Some optimization methods, such as SnobFit, benefit greatly from having the
search space reduced, because it alleviates scaling issues.
When possible, such bounds should be provided from the domain science.
When bounds derived from first principles are unavailable, an automatic way
of finding tighter bounds can be had by running a composition of optimizers.
To illustrate this principle we show the effect of optimizer composition by
using ImFil to derive tight bounds for SnobFit.

ImFil uses progressively smaller stencils in its search for the global
minimum (see Section~\ref{ssec:imfil}).
Once close enough, the combination of high noise levels and shallow
minimum means that no further progress can be made on the stencil, which
ImFil then labels as ``failed.''
The last good stencil provides the necessary bounds for SnobFit to proceed
and find a robust minimum.
The results of this approach are shown in Figure~\ref{fig:hubb_results}
(right) for Hubbard simulations with occupancies of 2 and 4 electrons.
In all cases, ImFil already outperforms the other optimizers, but SnobFit is
still able to improve from the point where ImFil fails.
Crucially, ImFil fails much earlier when noise levels are high
(see Section~\ref{sec:perf_considerations}), allowing the combined run of
ImFil+SnobFit to stay within budget.

\subsection{Performance Considerations}
\label{sec:perf_considerations}

Optimizer quality is also quantified by the total execution time.
We note that the wall time is completely dominated\footnote{The true ratio
depends on the quantum hardware chosen and the server CPU running the
classical optimizer. We estimate time spent in the classical step to be about
1\% of the total in typical setups.}
by the quantum chip for current devices.
When considering the optimizer in isolation, the number of objective function
evaluations is thus a good proxy for wall clock performance.

Most optimizers provide control over the number of evaluations per iteration,
thus determining single iteration overhead.
A certain minimum number of evaluations is always necessary to fill out a
stencil, apply a local model, or map a trust region.
The incremental improvement from adding more points to the current
iteration is, however, less than the improvement obtained from spending that
budget on an extra iteration.

Convergence criteria provide control over the total number of iterations.
Most optimizers define convergence as improvement between consecutive steps
falling below a threshold, or failing altogether a given number of times.
The lack of local improvement need not stop the search, e.g.\ for NOMAD and
SnobFit it can be chosen to initiate more {\em global} searches.
Whether those global searches are useful depends on the quality of the
initial and on the presence of local minima.

The details of the science problem matters greatly as well: tighter bounds
and a higher quality initial reduce the number of iterations needed
(see e.g.\ Figure~\ref{fig:search}).
An efficient ansatz with fewer parameters, for example through exploitation
of symmetries, and an optimization surface with steep gradients near the
global minimum, can also have a big impact.

Finally, there are differences intrinsic to the optimization methods.
Figure~\ref{fig:iterations} shows the number of evaluations for increasing
levels of noise, for both the ethylene rotation simulation (left) and the
Hubbard model with 4 electrons (right).
There is little sensitivity to noise in the (much simpler) rotation
simulation, except for BFGS which falls apart at high noise levels.
A clearer picture emerges in the Hubbard simulation: convergence criteria
that take into account the observed level of noise in their convergence
criteria work best.
E.g.\ PyBobyqa, which uses a fixed threshold, fails to converge, because
noise causes sufficient differences between iterations to remain above
threshold, leading it to use up the full budget.
The other optimizers, which either track overall improvement or improvement
within an iteration given the noise, stop  much earlier as noise
increases.
This is especially beneficial when conserving budget is important to allow
switching of optimizers, e.g.\ from ImFil to SnobFit as shown in the
previous section, while remaining within the budget overall.

\section{Discussion}
\label{sec:discusion}

A lot of research is spent on improving VQE quantum circuits to demonstrate
science results on NISQ hardware.
The need for noise-aware minimizers has been acknowledged, but its magnitude
may have been understated.
Our study indicates that using a classical optimizer that is not noise-aware
would make it the weakest link in the VQE chain: {\it use of specialized
noise robust optimizers is essential on NISQ hardware}.

Our evaluations further indicate that:
\begin{itemize}[leftmargin=*]
\item For noise-free optimization, SciPy optimizers such as BFGS or Cobyla
  are fastest by far.
  They do fail in the presence of even small noise, to the point of becoming
  unusable.
\item When decent parameter bounds are available, ImFil is preferable,
  followed by NOMAD.
  When tight bounds are available, SnobFit should be considered.
  A composition of optimizers works best for final solution quality.
\item When high quality initial parameters are available, trust region
  methods such as PyBobyqa are fastest and preferable, followed by NOMAD and
  to a lesser extent SnobFit.
  ImFil is not very sensitive to the value of the initial solution.
\item Taking performance data into account does not change the above
  recommendations.
  We do note that some optimizers are adaptive and properly reduce the number
  of evaluations in the presence of noise, e.g.\ ImFil and NOMAD.
\item To balance solution quality, execution time, and premature convergence,
  ImFil provides direct control over scales and searches.
  For the others, only limited control is possible by tweaking the
  convergence criteria, (attenuated) step sizes, points in the local model,
  and/or overall budget.
\end{itemize}

There are strong convergence requirements on the minimizer in terms of
distance to the global minimum~\cite{McClean2015}, but also constraints on the
number of evaluations before convergence as e.g.\ calibrations may drift.
The optimizer may need to find gradients on a surface with many local minima
due to the noise, and do so with the least number of iterations possible.
The ansatz in VQE directly drives the optimization surface and influences the
noise (e.g.\ through circuit depth), thus our study provides important
feedback for practical ansatz design.

Our results support the following conjectures:
\begin{itemize}[leftmargin=*]
\item There is no free lunch: a {\em suite} of minimizers is needed to match
  specific strengths to specific problems, making use of any available domain
  science information.
\item On-chip noise affects the objective non-linearly.
  Even if a global minimum value can not be found, a search for a
  {\em robust} minimum can still yield the correct optimal parameters.
\item For complex surfaces, noise can trap the optimizer in a local minimum.
  Domain science information is then needed to provide more or better
  constraints.
\item Most of the methods can scale up to hundreds of parameters.
  On NISQ hardware, with the minimizers provided, we expect the performance
  of hybrid approaches to be limited by the quantum part of the algorithms.
\end{itemize}

Overall, this study indicates that the success of VQE on NISQ devices is
contingent on the availability of classical optimizers that handle noisy
outputs well at the needed scales.
As of yet, this is an open research area, where our study details some of the
challenges to be expected.
Our software optimizers toolkit is directly useful to VQE Quantum Information
Science practitioners, as well as a good starting point for mathematicians in
search of better optimization methods tailored to VQE and other hybrid
quantum-classical algorithms.

\section{Related Work}
\label{sec:related}

For VQE, an initial discussion about optimization challenges in the
presence of noise is provided by McClean et al.~\cite{McClean2015}.
They study a UCC wavefunction for $H_2$, encoded into 4 qubits and optimize
over a single parameter.
Simulated measurement estimator noise is added to the objective function at
a specified variance ${\epsilon }^{2}$.
They compare Nelder-Mead with GLCLUSTER, LGO, and MULTIMIN (from TOMLAB,
motivated by the study by Rios et al.~\cite{Rios2013}).
Even for this single parameter problem, these optimizers face challenges
with noisy output.
Current QAOA~\cite{zhou2018quantum} studies still use BFGS and Nelder-Mead,
as they still concentrate mostly on the quantum algorithm part of the
problem.
While the VQE results are subject to physical or chemical laws which
bound their range, this does not apply to most QAOA approaches.
Thus, it is our expectation they will need to be supplemented with
optimizers robust in the presence of noise.

An orthogonal approach is the incorporation of error mitigation techniques.
The proposed zero-noise extrapolation techniques~\cite{Li_2017,Temme_2017}
seem to impose no constraints on optimizers and just run in the first
step the full VQE algorithm. An additional step calibrates the impact
of system noise, followed by an offline procedure to extrapolate
results to the ideal regimen of zero-noise. While the IBM
studies~\cite{Li_2017,Temme_2017} insert noise at the pulse level,
Dumitrescu et al.~\cite{Dumitrescu_2018} insert noise using additional
CNOT gates and describe a zero-noise extrapolation procedure.
Current results are for small circuits with few parameters (two) involved in
the optimization.
It remains to be seen whether they relax the requirements on robust
optimizers when applied to higher dimensional problems on more complex
optimization surfaces.

\comment{Another area of interest is the work in the numerical optimization
realm. Rios et al.~\cite{Rios2013} provide a comprehensive evaluation
of derivative-free numerical optimizers along multiple dimensions including
scalability and quality of solution, for convex and non-convex, smooth
and non-smooth surfaces. Overall, they recommend the
commercial TOMLAB~\cite{tomlab} implementations of GLCLUSTER, LGO and
MULTIMIN. Each is best for a given combination of surface convexity
and smoothness. Also note that all the  algorithms included in {\sc
  scikit-quant} are very close to any of the TOMLAB implementations
for some type of surface. }

\section{Conclusion}
\label{sec:conclusion}

Successful application of hybrid quantum-classical algorithms, involving a
classical optimizer, on current NISQ hardware, requires the optimizer to be
noise-aware.
We have collected a suite of optimizers in {\sc scikit-quant} that we found
to work particularly well, easily outperforming optimizers available through
the widely used standard SciPy software.

We studied VQE, but expect the results to be generally applicable:
by providing a suite of optimizers with consistent programming interfaces,
it is possible to easily apply combinations of optimizers, playing into
their respective strengths.
Our studies indicate that with these optimizers, the classical step is no
longer the weakest link on NISQ-era hardware.

\section*{Acknowledgment}
This work was supported by the DOE under contract DE-5AC02-05CH11231,
through the Office of Advanced Scientific Computing Research (ASCR) Quantum
Algorithms Team and Accelerated Research in Quantum Computing programs.

{\footnotesize
\bibliographystyle{abbrv}
\bibliography{minimizers,quantum,quant_chem}

\begin{thebibliography}{10}

\bibitem{qiskit}
{Abraham, H. {\em et.~al.}}
\newblock Qiskit: An open-source framework for quantum computing, 2019.

\bibitem{abramson2009}
M.~{Abramson}, C.~{Audet}, J.~{Dennis, Jr.}, and S.~{Le Digabel}.
\newblock Orthomads: a deterministic mads instance with orthogonal directions.
\newblock {\em SIAM Journal on Optimization}, 20(2):948--966, 2009.

\bibitem{audet2006}
C.~{Audet} and J.~{Dennis, Jr.}
\newblock Mesh adaptive direct search algorithms for constrained optimization.
\newblock {\em SIAM Journal on Optimization}, 17(1):188--217, 2006.

\bibitem{audet2009}
C.~{Audet} and J.~{Dennis, Jr.}
\newblock A progressive barrier for derivative-free nonlinear programming.
\newblock {\em SIAM Journal on Optimization}, 20(1):445--472, 2009.

\bibitem{pybobyqa1}
C.~{Cartis}, J.~{Fiala}, B.~{Marteau}, and L.~{Roberts}.
\newblock Improving the flexibility and robustness of model-based
  derivative-free optimization solvers.
\newblock Technical report, University of Oxford, 2018.

\bibitem{pybobyqa2}
C.~{Cartis}, L.~{Roberts}, and O.~{Sheridan-Methven}.
\newblock Escaping local minima with derivative-free methods: a numerical
  investigation.
\newblock Technical report, University of Oxford, 2018.

\bibitem{rbfopt}
A.~Costa and G.~Nannicini.
\newblock Rbfopt: an open-source library for black-box optimization with costly
  function evaluations.
\newblock {\em Mathematical Programming Computation}, 10(4):597--629, Dec 2018.

\bibitem{crooks2019gradients}
G.~E. Crooks.
\newblock Gradients of parameterized quantum gates using the parameter-shift
  rule and gate decomposition, 2019.

\bibitem{Dumitrescu_2018}
E.~Dumitrescu, A.~McCaskey, G.~Hagen, G.~Jansen, T.~Morris, T.~Papenbrock,
  R.~Pooser, D.~Dean, and P.~Lougovski.
\newblock Cloud quantum computing of an atomic nucleus.
\newblock {\em Physical Review Letters}, 120(21), May 2018.

\bibitem{farhi2014quantum}
E.~Farhi, J.~Goldstone, and S.~Gutmann.
\newblock A quantum approximate optimization algorithm, 2014.

\bibitem{feynman1982}
R.~P. {Feynman}.
\newblock {Simulating Physics with Computers}.
\newblock {\em International Journal of Theoretical Physics}, 21:467--488, June
  1982.

\bibitem{nelder-mead}
F.~Gao and L.~Han.
\newblock Implementing the nelder-mead simplex algorithm with adaptive
  parameters.
\newblock {\em Computational Optimization and Applications}, 51(1):259--277,
  Jan 2012.

\bibitem{cirq}
Google.
\newblock Cirq.
\newblock Available at https://github.com/quantumlib/Cirq.

\bibitem{cmapy}
N.~Hansen.
\newblock Cma-es, covariance matrix adaptation evolution strategy for
  non-linear numerical optimization in python.
\newblock Available at https://pypi.org/project/cma/.

\bibitem{cmaes}
N.~Hansen.
\newblock The {CMA} evolution strategy: {A} tutorial.
\newblock {\em CoRR}, abs/1604.00772, 2016.

\bibitem{snobfit}
W.~Huyer and A.~Neumaier.
\newblock Snobfit - {S}table {N}oisy {O}ptimization by {B}ranch and {F}it.
\newblock {\em ACM Trans. Math. Software}, 35(9), 2008.

\bibitem{ibmnature}
A.~Kandala, A.~Mezzacapo, K.~Temme, M.~Takita, M.~Brink, J.~M. Chow, and J.~M.
  Gambetta.
\newblock Hardware-efficient variational quantum eigensolver for small
  molecules and quantum magnets.
\newblock {\em Nature}, 549:242 EP --, 09 2017.

\bibitem{ImFil11}
C.~Kelley.
\newblock {\em Implicit Filtering}.
\newblock SIAM, 2011.

\bibitem{Kubler2020adaptiveoptimizer}
J.~M. K{\"{u}}bler, A.~Arrasmith, L.~Cincio, and P.~J. Coles.
\newblock An {A}daptive {O}ptimizer for {M}easurement-{F}rugal {V}ariational
  {A}lgorithms.
\newblock {\em {Quantum}}, 4:263, May 2020.

\bibitem{nomad}
S.~{Le Digabel}.
\newblock Nomad: Nonlinear optimization with the mads algorithm.
\newblock {\em ACM Trans. Math. Softw.}, 37(4):44:1--44:15, Feb. 2011.

\bibitem{Li_2017}
Y.~Li and S.~C. Benjamin.
\newblock Efficient variational quantum simulator incorporating active error
  minimization.
\newblock {\em Physical Review X}, 7(2), Jun 2017.

\bibitem{RevModPhys.92.015003}
S.~McArdle, S.~Endo, A.~Aspuru-Guzik, S.~C. Benjamin, and X.~Yuan.
\newblock Quantum computational chemistry.
\newblock {\em Rev. Mod. Phys.}, 92:015003, Mar 2020.

\bibitem{McClean_2020}
J.~R. McClean, Z.~Jiang, N.~C. Rubin, R.~Babbush, and H.~Neven.
\newblock Decoding quantum errors with subspace expansions.
\newblock {\em Nature Communications}, 11(1), Jan 2020.

\bibitem{McClean2015}
J.~R. McClean, J.~Romero, R.~Babbush, and A.~Aspuru-Guzik.
\newblock {The theory of variational hybrid quantum-classical algorithms}.
\newblock {\em New Journal of Physics}, 18(2):23023, 2016.

\bibitem{openfermion}
{McClean, J. {\em et.~al.}}
\newblock Openfermion: The electronic structure package for quantum computers,
  2017.

\bibitem{NoceWrig06}
J.~Nocedal and S.~J. Wright.
\newblock {\em Numerical Optimization}.
\newblock Springer, New York, NY, USA, second edition, 2006.

\bibitem{bobyqa}
M.~Powell.
\newblock The {BOBYQA} algorithm for bound constrained optimization without
  derivatives.
\newblock Report DAMTP 2009/NA06., Department of Applied Mathematics and
  Theoretical Physics, Cambridge University, 2009.

\bibitem{cobyla}
M.~J.~D. Powell.
\newblock {\em A Direct Search Optimization Method That Models the Objective
  and Constraint Functions by Linear Interpolation}, pages 51--67.
\newblock Springer Netherlands, Dordrecht, 1994.

\bibitem{Powell2014}
W.~Powell.
\newblock Clearing the jungle of stochastic optimization.
\newblock {\em INFORMS Tutorials in Operations Research: Bridging Data and
  Decisions}, Published online: 27 Oct 2014:109--137, 2014.

\bibitem{dycors}
R.~{Regis} and C.~{Shoemaker}.
\newblock Combining radial basis function surrogates and dynamiccoordinate
  search in high-dimensional expensive black-box optimization.
\newblock {\em Engineering Optimization}, 45(5):529--555, 2013.

\bibitem{Rios2013}
L.~M. Rios and N.~V. Sahinidis.
\newblock Derivative-free optimization: a review of algorithms and comparison
  of software implementations.
\newblock {\em Journal of Global Optimization}, 56(3):1247--1293, Jul 2013.

\bibitem{urlsciq}
Scikit-quant.
\newblock \url{scikit-quant.org}.

\bibitem{scipy_web}
Scipy documentation.
\newblock \url{https://www.scipy.org}.

\bibitem{projq}
S.~Steiger, T.~{H\"{a}ner}, and M.~Troyer.
\newblock {ProjectQ: An Open Source Software Framework for Quantum Computing}.
\newblock {\em ArXiv e-prints}, Dec. 2016.

\bibitem{Temme_2017}
K.~Temme, S.~Bravyi, and J.~M. Gambetta.
\newblock Error mitigation for short-depth quantum circuits.
\newblock {\em Physical Review Letters}, 119(18), Nov 2017.

\bibitem{PhysRevA.94.052325}
J.~J. Wallman and J.~Emerson.
\newblock Noise tailoring for scalable quantum computation via randomized
  compiling.
\newblock {\em Phys. Rev. A}, 94:052325, Nov 2016.

\bibitem{Wang_2018}
Z.~Wang, S.~Hadfield, Z.~Jiang, and E.~G. Rieffel.
\newblock Quantum approximate optimization algorithm for maxcut: A fermionic
  view.
\newblock {\em Physical Review A}, 97(2), Feb 2018.

\bibitem{zhou2018quantum}
L.~Zhou, S.-T. Wang, S.~Choi, H.~Pichler, and M.~D. Lukin.
\newblock Quantum approximate optimization algorithm: Performance, mechanism,
  and implementation on near-term devices, 2018.

\end{thebibliography}
}

\end{document}